\documentclass[twocolumn,showpacs,nofootinbib,preprintnumbers,amsmath,amssymb,showkeys]{revtex4}

\usepackage{epsfig}
\usepackage{color}% Include figure files
\usepackage{graphicx}% Include figure files
\usepackage{dcolumn}% Align table columns on decimal point
\usepackage{bm}% bold math
\usepackage{dsfont}
\newcommand{\beq}{\begin{equation}}
\newcommand{\eeq}{\end{equation}}
\newcommand{\bea}{\begin{eqnarray}}
\newcommand{\eea}{\end{eqnarray}}
\newcommand{\nn}{\nonumber \\}

\newcommand\eqn[1]{(\ref{#1})}      % parentheses around the LaTex "ref" macro
\newcommand\Eqn[1]{Eq.~(\ref{#1})}  % includes ``Eq.'' in front
\newcommand\Fig[1]{Fig.~\ref{#1}}  % includes ``Eq.'' in front

\newcommand{\tr}{\hbox{tr}}

\newcommand{\cb}{{\bar c}}

\begin{document}

%\preprint{APS/123-QED}

\title{Deconfinement transition in SU($2$) Yang-Mills theory: A two-loop study
}% Force line breaks with \\

\author{U. Reinosa}%
\affiliation{%
Centre de Physique Th\'eorique, Ecole Polytechnique, CNRS, 91128 Palaiseau Cedex, France
}%
\author{J. Serreau}%
\affiliation{%
 Astro-Particule et Cosmologie (APC), CNRS UMR 7164, Universit\'e Paris 7 -- Denis Diderot\\ 10, rue Alice Domon et L\'eonie Duquet, 75205 Paris Cedex 13, France
}%
\author{M. Tissier}
\affiliation{LPTMC, Laboratoire de Physique Th\'eorique de la Mati\`ere Condens\'ee\\ CNRS UMR 7600, Universit\'e Pierre et Marie Curie, \\ boite 121, 4 place Jussieu, 75252 Paris Cedex 05, France
}
\author{N. Wschebor}%
\affiliation{%
 Instituto de F\'{\i}sica, Facultad de Ingenier\'{\i}a, Universidad de la Rep\'ublica \\ J.H.y Reissig 565, 11000 Montevideo, Uruguay
}%

\date{\today}% It is always \today, today,
             %  but any date may be explicitly specified

\begin{abstract}
In a recent work we have proposed a perturbative approach for the study of the phase transition of pure Yang-Mills theories at finite temperature. This is based on a simple massive extension of background field methods in the Landau-DeWitt gauge, where the gluon mass term is related to the existence of Gribov ambiguities. We have shown that a one-loop calculation of the background field effective potential describes well the phase structure of the SU($2$) and SU($3$) theories. Here, we present the calculation of the next-to-leading-order contribution in perturbation theory for the SU($2$) case. In particular, we compute the background field effective potential at two-loop order and the corresponding Polyakov loop, a gauge invariant order parameter of the transition, at one-loop order. We show that the two-loop correction brings the critical temperature closer to its actual value as compared to the previous one-loop result. We also compute the thermodynamic pressure as a function of the temperature and show that two-loop contributions play an important role in the vicinity of the phase transition.
 \end{abstract}

% PACS, the Physics and Astronomy
                             % Classification Scheme.
%Use showkeys class option if keyword
                              %display desired
\maketitle

%\tableofcontents

%%%%%

%%%%%   Introduction
\section{Introduction}
\label{sec:intro}

The deconfinement transition of hadronic matter into a plasma of quarks and gluons at high temperature is a remarkable phenomenon. It is thought to have played a role in the early Universe and it is the major subject of investigation of ultrarelativistic heavy ion collisions at CERN and at RHIC. Establishing firmly the existence of this transition in QCD and characterizing its properties is a formidable task which has only been possible thanks to more than three decades of dedicated lattice studies \cite{Engels:1980ty,McLerran:1980pk,Kuti:1980gh}; see Refs.~\cite{Petreczky:2012rq,Philipsen:2012nu} for recent reviews. Early calculations clearly established the existence of a phase transition in pure SU($N$) Yang-Mills theories, related to the spontaneous breaking of the center ($Z_N$) of the gauge group \cite{Engels:1981qx,Boyd:1995zg}. The inclusion of dynamical quarks, first with heavy masses and then with physical masses, has been a central issue in the field for years. Only recently have precise calculations of the thermodynamic properties of QCD with physical quark masses been achieved, with the result that the phase transition of the pure gauge theory becomes a crossover in QCD \cite{Bazavov:2014pvz,Borsanyi:2013bia}. 

Thanks to asymptotic freedom, standard perturbative approaches make sense at high temperatures and an intense activity has been concerned with computing the thermodynamics and transport properties of the quark-gluon plasma by means of (semi)analytical methods \cite{Blaizot:1999ip,Blaizot:2000fc,Blaizot:2001vr,Andersen:1999sf,Andersen:2002ey,Andersen:2011sf,Haque:2013sja,Arnold:2000dr,Arnold:2003zc}. In this regime, infrared divergences call for the resummation of infinite subclasses of perturbative diagrams, the so-called hard thermal loops \cite{Braaten:1989mz,Braaten:1990az,Braaten:1991gm}. Such high temperature approaches reproduce the thermodynamic properties of the deconfined plasma down to a few times the transition temperature \cite{Blaizot:1999ip,Blaizot:2000fc,Blaizot:2001vr} but fail to capture the physics of the phase transition. It is commonly accepted that the low temperature confining phase cannot be described by means of perturbation theory because of the existence of a Landau pole at low energy, where the running coupling diverges. 

Existing nonperturbative continuum descriptions of the transition region are based on truncations of Dyson-Schwinger equations (DSE), nonperturbative/functional renormalization-group techniques \cite{Marhauser:2008fz,Braun:2007bx,Braun:2010cy,Fister:2013bh,Epple:2006hv,Alkofer:2006gz,Fischer:2009wc,Fischer:2009gk}, the Hamiltonian approach of \cite{Reinhardt:2012qe,Reinhardt:2013iia}, or two-particle-irreducible (2PI) inspired approaches \cite{Fukushima:2012qa,Fister:2013bh}. These have the advantage over lattice calculations that they can easily be used at finite chemical potential \cite{Fischer:2013eca,Fischer:2014vxa,Fischer:2011mz,Fischer:2012vc} and/or for computing real-time quantities \cite{Mueller:2010ah,Strauss:2012dg,Haas:2013hpa}. To be trustable in these situations though, they have to be tested against lattice calculations in situations where the latter is well under control. Functional renormalization-group (FRG) methods have been shown to correctly describe the phase structure of pure gauge theories, with transition temperatures in agreement with lattice results \cite{Braun:2007bx,Fister:2013bh}. Such approaches also provide nontrivial insight concerning dynamical aspects of the deconfinement transition. For instance, an interesting connection between the confinement of static quarks and the infrared (IR) behavior of gluon and ghost correlators has been pointed out \cite{Braun:2007bx,Fister:2013bh}. Of course, neither the DSE nor the FRG can be solved exactly and one's ability to perform actual calculations relies on several assumptions which, even when these are well motivated, can be difficult to check explicitly. A general criticism that many nonperturbative approaches have to face is that they do not always involve a systematic approximation scheme and it is often difficult to compute corrections to the obtained results. Still, the DSE/FRG provide the most powerful nonperturbative tools---apart from lattice techniques---to investigate the physics of the deconfinement transition directly at the level of the basic degrees of freedom of the theory.

In a series of recent works \cite{Tissier_10,Tissier_11,Serreau:2012cg,Pelaez:2013cpa,Serreau:2013ila,Reinosa:2013twa,Pelaez:2014mxa,Reinosa:2014ooa}, yet a different route for the study of the infrared dynamics of Yang-Mills fields has been proposed. This is based on simple massive extensions of the standard Faddeev-Popov Lagrangian in the Landau gauge and in the Landau-DeWitt gauge.\footnote{A particular class of nonlinear covariant gauges has been considered along similar lines in \cite{Serreau:2013ila}.} This is motivated by the observation that perturbative calculations of Yang-Mills correlators in the vacuum \cite{Tissier_10,Tissier_11,Pelaez:2013cpa,Pelaez:2014mxa} and at finite temperature \cite{Reinosa:2013twa} in the massive extension of the Landau gauge action\footnote{This is a particular case of the Curci-Ferrari Lagrangian \cite{Curci76}.} agree well with lattice data down to deep IR momenta. It is worth emphasizing that, if the typical value of the gauge coupling $g$ required for such comparisons is of a few units, the relevant expansion parameter at zero temperature is $g^2N/(16\pi^2)\lesssim1$. An important feature of the massive theory is that there exist IR safe renormalization-group trajectories, with no Landau pole \cite{Tissier_11,Serreau:2012cg}. Moreover, it has been argued in \cite{Serreau:2012cg} that such a massive extension naturally arises as an effective theory for Yang-Mills correlators in a new one-parameter family of Landau gauges which, unlike the standard Faddeev-Popov construction, takes into account the existence of Gribov ambiguities. In this approach, the gluon mass term appears as a gauge-fixing parameter which lifts the degeneracy between Gribov copies. 

This has been extended to the Landau-DeWitt gauge in the context of background field methods in \cite{Reinosa:2014ooa}. There, we have shown that a calculation of the background field potential at first nontrivial (one-loop) order in perturbation theory correctly reproduces the phase structure of SU($N$) theories: one finds a confining phase at low temperature and a transition to a deconfined phase at high temperature which is second order for $N=2$ and first order for $N=3$, with transition temperatures in qualitative agreement with known values from lattice calculations. Our one-loop results are similar to those of the FRG studies of \cite{Braun:2007bx} and actually corroborate the related findings concerning the relation between the IR behavior of gluon and ghost propagators and the existence of a confining phase at low temperature. Definite advantages of such a perturbative approach are, first, that low-order calculations are technically very simple and, second, that they can be systematically improved by computing higher orders. If at asymptotically high temperatures, the expansion parameter is $g$ due to collective infrared effects which necessitate the resummation of hard thermal loops \cite{Blaizot:2001nr}, near the transition region and below, the effective gluon mass tames some of the infrared problems of the perturbative series and it is not clear what is the relevant expansion parameter. It is the purpose of the present article to study the importance of such higher-order terms by computing the background field potential, the Polyakov loop and the thermodynamic pressure at next-to-leading order in the perturbative expansion.

Before embarking in actual calculations, let us make some general comments concerning the massive extension of the Faddeev-Popov action in the class of (Landau or Landau-DeWitt) gauges considered here. What usually prevents a mass term in the gauged-fixed action is obviously not gauge invariance, but BRST symmetry. The latter is a property of the Faddeev-Popov action, which is known to be valid at best in the high energy perturbative regime. But the Faddeev-Popov construction is certainly wrong in general since it ignores the existence of Gribov copies and, hence, does not completely fix the gauge. In fact, it is well known that the only truly nonperturbative formulation of the gauge-fixed theory known so far, that is the lattice, cannot accommodate the BRST symmetry without leading to undefined zero over zero ratios \cite{Neuberger:1986vv,Neuberger:1986xz}.

 A consistent quantization procedure, free of Gribov ambiguities, is likely to break the BRST symmetry. A well-known example is the minimal Landau gauge on the lattice, where one picks up a random Gribov copy on each gauge orbit \cite{Cucchieri:2014via}. Examples in the continuum include the so-called (refined) Gribov-Zwanziger approach \cite{Gribov77,Zwanziger89,Zwanziger92,Vandersickel:2012tz,Dudal08} or the averaging procedure of \cite{Serreau:2012cg}, already mentioned. In the latter case, the bare gluon mass originates from the averaging procedure and is simply a gauge-fixing parameter which explicitly breaks the BRST symmetry. Such gauge-fixing procedures provide efficient starting points for perturbative calculations of Yang-Mills correlators and, for the latter, of the phase structure of the theory at finite temperature. In both approaches the BRST breaking is soft and the gauge-fixed actions present modified (non-nilpotent) BRST symmetries, which ensure their renormalizability.\footnote{We stress, however,  that the breaking of the nilpotent BRST symmetry invalidates the standard proof of unitarity. The question whether such theories are (perturbatively) unitary is still an open problem.} At this point, it is worth emphasizing that the continuum approaches mentioned above also introduce, in one way or another, a BRST-breaking ingredient. This typically appears through choices of boundary conditions and/or ultraviolet subtractions in the context of DSE \cite{Dudal:2014rxa}, or through the infrared regulator in FRG approaches. 

The plan of the paper is as follows. Section \ref{sec:first} sets the scene and briefly recalls the basics of the $Z_N$ transition and of static quark confinement. In Sec.~\ref{sec:generalities} we present the massive extension of the Landau-DeWitt gauge and derive the corresponding Feynman rules. In Sec.~\ref{sec:potential}, we summarize the calculation of the two-loop correction to the background field effective potential and in Sec.~\ref{sec:polyakov_loop}, we give the corresponding expression for the one-loop correction to the Polyakov loop. Finally, in Sec.~\ref{sec:results}, we present our results concerning the order of the phase transition and the value of the transition temperature as well as the temperature dependence of the Polyakov loop and of the thermodynamic pressure. The essential steps of our calculations are presented in the main text while the technical details are gathered in the Appendixes. 

Although we specify to the case $N=2$ throughout this work, some formulas are valid for arbitrary $N$. An important observation is that many steps of our calculations are similar to those in the massive extension of the Landau gauge, i.e., at vanishing background field. In fact, when expressed in an appropriate color basis, the Feynman rules of the theory have the same form as those in the Landau gauge, with the only difference that the momenta get shifted by an amount proportional to the background field. The key point is that, because of the residual global color symmetry, these shifts are conserved at the interaction vertices. This allows us to use various simplifying manipulations, detailed in Appendix~\ref{appsec:gluloupe}, and to reduce all the two-loop diagrams contributing to the background field potential to simple scalarlike sum-integrals; see Appendix~\ref{appsec:reduc}. The evaluation of the corresponding Matsubara sums is presented in Appendix~\ref{appsec:matsubara}, which allows us to write the background field potential in a rather simple form in Appendix~\ref{appsec:final}. In Appendix~\ref{appsec:PL}, we detail the calculation of the Polyakov loop at one-loop order in the presence of the nontrivial background field. In particular, this demonstrates explicitly that the Polyakov loop vanishes if and only if the minimum of the background field potential takes particular, $Z_2$--symmetric values. This confirms, at this order of approximation, that the background field itself can be used as an order parameter for confinement, as advocated in \cite{Braun:2007bx,Braun:2010cy,Fister:2013bh}.\footnote{To our knowledge, this has only be proven explicitly for the SU($2$) theory in the Polyakov gauge in \cite{Marhauser:2008fz}.} Finally, we provide, in Appendix~\ref{appsec:proof}, a general proof that the Polyakov loop vanishes when the background field takes $Z_2$--symmetric values. We do not have a similar proof for the converse.

%%%%%   Confinement-deconfinement transition and Polyakov loop
\section{Confinement-deconfinement transition in Yang-Mills theory}
\label{sec:first}

We consider the Euclidean Yang-Mills action in $d=4-2\epsilon$ dimensions
\beq 
\label{eq:YM}
S_{\rm  YM}[A]=\frac{1}{2}\int_x \tr\left\{F_{\mu\nu}F_{\mu\nu}\right\}\,, 
\eeq 
where $F_{\mu\nu}=\partial_\mu A_\nu-\partial_\nu
A_\mu -ig_0[A_\mu,A_\nu]$, with $g_0$ the bare coupling constant and $A$ the bare gauge field, $iA$ being an element of the Lie algebra of SU($N$). We have also defined $\int_x\equiv\int_0^\beta d\tau\int d^{d-1} x$, with $\beta=1/T$ the inverse temperature.

Let us recall some basic aspects of the deconfinement transition to be considered below. The free energy $F_q$ of an isolated static quark in a thermal gluon bath is directly related to the expectation value of the traced Polyakov loop---which we refer to as the Polyakov loop for short in what follows---as \cite{Svetitsky:1985ye}
\beq\label{eq:f_pol}
e^{-\beta (F_q-F_0)}=\frac{1}{N}\tr\left< P\exp\left\{ig_0\int_0^\beta \!\!d\tau A_0(\tau,{\bf x})\right\}\right>\equiv \ell,
\eeq 
where $F_0$ is the free energy in the absence of quark. Here, $P$ orders the matrix fields $A_0(\tau,{\bf x})$ from left to right with decreasing value of their time arguments and the brackets refer to the average in the theory defined by the action (\ref{eq:YM}). A vanishing $\ell$ signals an infinite free energy, hence a quark confining phase, while $\ell\neq0$ is interpreted as a phase where isolated static quarks are energetically allowed. 

It is well known \cite{Svetitsky:1985ye} that the Polyakov loop gets multiplied by a phase under generalized gauge transformations that leave the Yang-Mills action at finite temperature invariant and which are $\beta$ periodic in imaginary time, up to an element of the center of the group. This means that the deconfined phase is necessarily a phase where this symmetry group, or more precisely the quotient\footnote{One needs to consider the quotient group because the standard gauge transformations leave the Polyakov loop unchanged and thus a nonvanishing Polyakov loop does not tell anything about this subgroup. It is easily checked that the group of standard gauge transformations forms a normal subgroup within the group of generalized gauge transformations and thus the quotient group inherits a group structure, isomorphic to $Z_N$.} of the group of generalized gauge transformations by the subgroup of standard gauge transformations---which is isomorphic to $Z_N$---is spontaneously broken. Note that the converse is not necessarily true: Although this is not the expected behavior, one could, in principle, imagine a situation where the center is spontaneously broken but where the Polyakov loop still vanishes, the breaking being only manifest at the level of higher-order correlations. This emphasizes the fact that the confined or deconfined nature of the system, in the sense described above, is not the breaking of the $Z_N$ symmetry itself but really the zero or nonzero value of the Polyakov loop (or of any equivalent order parameter). In what follows, we shall compute the Polyakov loop at one-loop order within the massive extension of the Landau-DeWitt gauge put forward in~\cite{Reinosa:2014ooa}. 

%%%%%   The massive Landau-DeWitt action
\section{The massive Landau-DeWitt action}\label{sec:generalities}

%%%   General set-up
\subsection{General setup}
\label{sec:IIIA}

We quantize the theory using the background field method \cite{DeWitt:1967ub,Abbott:1980hw,Abbott:1981ke,Weinberg:1996kr}, where we introduce an {\it a priori} arbitrary background field configuration  $\bar A_\mu$ and define the fluctuating field $a_\mu=A_\mu-\bar A_\mu$. The Landau-DeWitt gauge condition reads
\beq
\label{eq:LdW}
 \bar D_\mu a_\mu=0,
\eeq
where $\bar D_\mu\varphi=\partial_\mu\varphi -ig_0 [\bar A_\mu,\varphi]$ for any field $i\varphi$ in the Lie algebra of the gauge group. Our gauge-fixed action reads \cite{Reinosa:2014ooa}
\beq
\label{eq_gf}
 S=\!\int_x\tr\left\{{1\over2}F_{\mu\nu}F_{\mu\nu}+{m_0^2}a_\mu a_\mu+2\bar D_\mu\bar cD_\mu c+2ih\bar D_\mu a_\mu\right\}\!,
\eeq
with $h$ a (real) Nakanishi-Lautrup field and $c$ and $\bar c$ the Faddeev-Popov ghosts and antighost fields. In terms of the field $a_\mu$, we have
\begin{align}
\label{eq:Fieldstrength}
 F_{\mu\nu}&=\bar F_{\mu\nu}+\bar D_\mu a_\nu-\bar D_\nu a_\mu-ig_0 [a_\mu, a_\nu],
\end{align}
with $\bar F_{\mu\nu}^a=F_{\mu\nu}^a[\bar A]$ the field strength tensor evaluated at $A=\bar A$, and 
\beq
 D_\mu \varphi=\partial_\mu \varphi-ig_0 [A_\mu,\varphi]=\bar D_\mu\varphi-ig_0 [a_\mu,\varphi].
\eeq
The action \eqn{eq_gf} has the obvious property
\beq
\label{eq_ginvbare}
 S[\bar A, \varphi]=S[\bar A^U, U\varphi U^{-1}],
\eeq
where $\varphi=(a,c,\bar c,h)$, $U$ is a local SU($N$) matrix, and 
\begin{equation}
\label{eq:transfofo}
  \bar A_\mu^U=U\bar A_\mu U^{-1}+\frac i {g_0}U\partial_\mu U^{-1}.
\end{equation}
At the level of the (quantum) effective action $\Gamma$ this implies \cite{Weinberg:1996kr}
\beq
\label{eq_ginv}
 \Gamma[\bar A, \varphi]=\Gamma[\bar A^U, U\varphi U^{-1}],
\eeq
where the fields $\varphi$ are now to be understood as average values in the presence of sources.\footnote{In particular, $h$ is now purely imaginary.} 

In principle, to evaluate physical observables at zero sources, one should minimize $\Gamma[\bar A,\varphi]$ with respect to $\varphi$ at a given $\bar A$. It can be argued, however, that one can alternatively minimize the functional
\beq
\label{eq_functilde}
 \tilde\Gamma[\bar A]=\Gamma[\bar A,0]
\eeq
with respect to the background field $\bar A$ \cite{Braun:2007bx,Fister:2013bh,Reinosa:2014ooa}. This functional obeys the background gauge symmetry
\beq\label{eq:bgsym}
\tilde\Gamma[\bar A]=\tilde\Gamma[\bar A^U],
\eeq
which is trivially preserved in perturbation theory. The Polyakov loop can be obtained as\footnote{The (gauge invariant) Polyakov loop is independent of the background $\bar A$ and evaluating it at $\bar A=\bar A_{\rm min}$ is a matter of choice. It proves convenient though since $\langle a_0(x)\rangle_{\rm min}=0$, thus avoiding the calculation of $\langle a_0(x)\rangle$ in the presence of a generic $\bar A$.}
\beq
\label{eq:ploop}
 \ell(T)=\frac{1}{N}\tr\left< P\exp\left\{ig_0\int_0^\beta \!\!d\tau (\bar A_0+a_0)(x)\right\}\right>_{\rm min},
\eeq
with $x\equiv(\tau,{\bf x})$ and where the brackets stand for an average in the gauge-fixed theory (\ref{eq_gf}). The right-hand side of \Eqn{eq:ploop} is evaluated at an absolute minimum $\bar A(x)=\bar A_{\rm min}(x)$ of $\tilde\Gamma[\bar A]$. Because the Polyakov loop involves only the temporal component of the background field and because $\bar A_{\rm min}(x)=\langle A(x)\rangle_{\rm min}$ (since $\langle a(x)\rangle_{\rm min}=0$ by construction) is $x$ independent, it is sufficient to consider homogeneous background fields in the temporal direction $\bar A_\mu(x)=\bar A_0\delta_{\mu0}$. Moreover, the Hermitian matrix $\bar A_0$ can be diagonalized by means of a global SU($N$) rotation and one can thus, with no loss of generality, restrict $\bar A_0$ to the Cartan subalgebra of the color group. We shall write $\bar A_0=\bar A_0^k t^k$, where $t^k$ are the SU($N$) generators in the Cartan subalgebra.  We thus have to minimize the background field effective potential
\beq
\label{eq:poteff}
 V(T,r^k)=\frac{\tilde\Gamma[\bar A]}{\beta \Omega}-V_{\rm vac},
\eeq
 where $r^k= g_0\beta\bar A_0^k$ and $\Omega$ is the spatial volume. We have subtracted $V_{\rm vac}$, the value of the potential at zero temperature which is independent of $r^\kappa$, as we shall see below. Finally, the thermodynamic pressure is simply given by 
\beq
\label{eq:pressure}
 p(T)=-V(T,r_{\rm min}^k(T))\,,
\eeq
where  $r_{\rm min}^k(T)= g_0\beta\bar A^k_{0,{\rm min}}(T)$.

 The symmetry \eqn{eq:bgsym} implies that the potential \eqn{eq:poteff} is invariant under gauge transformations of the form 
\beq
 U(\tau,{\bf x})=\exp \{i\tau\varphi/\beta\},
\eeq
where $\varphi$ is such that $U^{-1}(0,{\bf x})U(\beta,{\bf x})\propto\mathds{1}$. This implies that the potential is periodic along certain directions in the Cartan subalgebra.  Together with the invariance under global color rotations and charge conjugation,\footnote{In the SU($2$) case, charge conjugation of the gluon field can be seen as a global color rotation. This is not the case for SU($3$).} this implies that some of the extrema of the potential have specific locations \cite{Reinosa:2014ooa}. In the SU($2$) case, these considerations show that the potential is $2\pi$--periodic in $r$ and has extremas at $r=0\,({\rm mod}\,\pi)$. In Appendix  \ref{appsec:proof}, we provide a general proof that, among those, the values $r=\pi\,({\rm mod}\,2\pi)$ are such that $\ell=0$ and thus correspond to a confined phase. This supports the general expectation \cite{Braun:2007bx,Braun:2010cy,Fister:2013bh} that the background field itself can be used as an order parameter. Our proof is, however, not complete because the converse, i.e., $\ell=0\Rightarrow r=\pi({\rm mod}\,2\pi)$, is missing. In the present work, we shall, however, explicitly show that the equivalence holds at next-to-leading order in the loop expansion.

%%%   Renormalization
\subsection{Renormalization}

We introduce renormalized parameters and fields, related to the corresponding bare quantities in the usual way: 
\beq
 m_0^2=Z_{m^2}m^2\,,\quad g_0=Z_g g\,,
\eeq
and
\beq
\label{eq:renormfactors}
\begin{tabular}{ccccccc}
 ${\bar A}$&$\to$& $\sqrt{Z_{\bar A}}\,\bar A$&$,$&$\quad a$&$\to$& $\sqrt{Z_a}\,a\,,$\\
 \\
 $c$&$\to$& $\sqrt{Z_c}\,c$&$,$&$\quad \bar c$&$\to$&$ \sqrt{Z_c}\,\bar c\,,$
 \end{tabular}
\eeq
where we have kept the same notation for bare (left) and renormalized (right)  fields for simplicity. Notice that the background field $\bar A$ and the fluctuating field $a$ have different renormalization factors \cite{Binosi:2013cea}. The background field gauge symmetry \eqn{eq_ginv} implies that the product $g_0\bar A$ is finite \cite{Weinberg:1996kr}. In the following we impose the renormalization condition 
\beq
\label{eq:condgA}
 Z_g\sqrt{Z_{\bar A}}=1
\eeq
for the finite parts as well, so that $g_0\bar A\to g\bar A$. From here on, we only consider renormalized quantities unless explicitly stated. The values of the parameters $m$ and $g$ must be fixed from some external input, e.g., lattice data.\footnote{In the gauge-fixing procedure proposed in \cite{Serreau:2012cg,Reinosa:2014ooa}, the bare parameter $m_0^2>0$ controls the degeneracy lift between Gribov copies along each gauge orbit. In the continuum limit, it has to be sent to zero together with the bare coupling $g_0$, keeping renormalized parameters $m^2$ and $g$ finite.}

To set the value of the renormalized mass $m$, one would ideally use the value of a physical observable such as a glueball mass. An easier possibility in practice is to fix this parameter by employing lattice results for gauge-dependent quantities such as the the Yang-Mills correlators. In principle, this requires lattice results in the same gauge as described above, involving an average of Gribov copies. Instead, existing gauge-fixed lattice calculations typically select a particular Gribov copy in the so-called first Gribov region, where the Faddeev-Popov operator is positive definite. Still, explicit calculations in the massive extension of the Faddeev-Popov Lagrangian in the Landau gauge show that there exists a value of the renormalized mass which allows one to quantitatively reproduce the lattice data for the Yang-Mills correlators at vanishing temperature \cite{Reinosa:2013twa}. 

No lattice calculation exists so far in the Landau-DeWitt gauge with homogeneous temporal background field as considered here.\footnote{Such calculations are, in principle, feasible---see \cite{Cucchieri:2012ii}---and would be of great interest.} However, at zero temperature the background field vanishes and the latter reduces to the standard Landau gauge. In the present work, we shall thus use the values of $m$ and $g$ inferred from lattice calculations in the Landau gauge at vanishing temperature. To be consistent with the approximation order considered here, we need the one-loop expressions of the vacuum propagators. These have been computed in \cite{Tissier_11}. Using the renormalization conditions
\beq\label{eq:ren_cond}
 \Sigma_{\rm vac}(K^2=\mu^2)=\Pi_{\rm vac}(K^2=0)=\Pi_{\rm vac}(K^2=\mu^2)=0,
\eeq
where $\Sigma_{\rm vac}(K^2)$ and $\Pi_{\rm vac}(K^2)$ denote, respectively, the renormalized ghost and transverse gluon self-energies in the vacuum, the best fits to lattice data give $m\simeq 680$~MeV and $g\simeq 7.5$, with $\mu=1$~GeV for SU($2$) in $d=4$.

%%%   Feynman rules
\subsection{Feynman rules}
\label{sec:Frules}

For the homogeneous background fields considered here, $\bar A_\mu(x)=\bar A_0\delta_{\mu0}$, the curvature term vanishes: $\bar F_{\mu\nu}=0$. Moreover, as emphasized above, the background field $\bar A_0$ can be taken in the Cartan subalgebra of the gauge group. For SU($2$), the latter has only one direction which we choose along the third axis in color space. It is convenient to work with the basis of generators
\beq
\label{eq:basis}
 t^0=t^3\,, \quad t^\pm=\frac{t^1\pm i t^2}{\sqrt{2}},
\eeq 
which satisfy 
\beq
 [t^\kappa,t^\lambda]=\varepsilon^{\kappa\lambda\tau}t^{-\tau}\,, \quad \tr\{t^\kappa t^{-\lambda}\}=\frac{\delta^{\kappa\lambda}}{2}
\eeq
where $\varepsilon^{\kappa\lambda\tau}$ is the completely antisymmetric tensor, with $\varepsilon^{0+-}=1$.
We denote any element of the gauge group Lie algebra as $\varphi=\varphi^\kappa t^{\kappa}$.\footnote{In terms of the components in the Cartesian basis, one has $\varphi^0=\varphi^3$ and $\varphi^\pm=(\varphi^1\mp i\varphi^2)/\sqrt{2}$.} The components of the background covariant derivative are
\beq
\label{eq:barderivative}
 \left(\bar D_\mu \varphi\right)^\kappa=\left(\partial_\mu-i\kappa g\bar A_\mu^3\right) \varphi^\kappa\to -iK_\mu^{\kappa}\, \tilde\varphi^\kappa(K),
\eeq
where our convention for the Fourier transform is  
\beq
 \tilde\varphi(K)=\int_x e^{iK\cdot x}\varphi(x),
\eeq
such that $\partial_\mu\to-iK_\mu$. Here, we defined the shifted momentum 
\beq
 K_\mu^\kappa=K_\mu+\kappa g\bar A^3_\mu,
\eeq 
which satisfies
\beq
\label{eq:idmom}
 (-K)_\mu^{-\kappa}=-K_\mu^{\kappa}.
\eeq

\begin{figure}[t!]  
\begin{center}
\epsfig{file=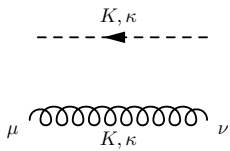,width=4cm}
 \caption{Diagrammatic representation of the ghost (dashed) and gluon (curly) propagators for momentum $K$ and color charge $\kappa$. The orientation of the flow of momentum and color charge is arbitrary.}\label{fig:propags}
\end{center}
\end{figure}

The background field breaks the global color group but there remains a residual symmetry under those color rotations that leave it invariant. In the SU($2$) theory, this is the group of SO($2$) $\sim$ U($1$) transformations, under which $\varphi^0\to\varphi^0$ and $\varphi^\pm\to e^{\pm i \alpha}\varphi^\pm$, with $\alpha$ a constant phase. Accordingly, we refer to $\varphi^0$ as the ``neutral'' component and to $\varphi^\pm$ as the ``charged'' components. The (temporal) background field plays the role of an external Abelian field coupled to the charged components, which leads to a simple shift of momentum\footnote{For a background field in the temporal direction, as considered here, only the (Matsubara) frequencies are shifted.} $\pm g\bar A_\mu^3$. Equivalently, it can be seen as an imaginary chemical potential associated with the conserved U($1$) charge. The residual color symmetry guarantees that the corresponding charge is conserved upon propagation and at the interaction vertices.

It is an easy exercise to compute the Feynman rules of the theory in the basis \eqn{eq:basis}. To each propagator and to each leg of an interaction vertex is associated a flow of color charge, which we define to follow the flow of momentum. The tree-level ghost and gluon propagators for momentum $K$ and charge state $\kappa$, represented in \Fig{fig:propags}, are given by 
\begin{align}
\label{eq:ghpropag}
 \langle c^{-\kappa}(-K) \cb^{\,\kappa}(K) \rangle&\equiv G^\kappa(K)=G_0(K^\kappa),\\
\label{eq:glupropag}
 \langle a_\mu^{-\kappa}(-K) a_\nu^{\kappa}(K) \rangle&\equiv G_{\mu\nu}^\kappa(K)={\cal P}_{\mu\nu}(K^\kappa)G_m(K^\kappa),
\end{align}
where we denote the scalar propagator of mass $\alpha$ by\footnote{Throughout this paper, we shall use the greek letters $\kappa$, $\lambda$, $\tau$, $\xi$, and $\omega$ to denote the various color (charge) states $0,+,-$, while $\mu$, $\nu$, $\rho$, and $\sigma$ are used for spacetime indices. In the Appendixes we also employ $\alpha$, $\beta,$ and $\gamma$ to denote various possible mass states.}
\beq
\label{eq:propalpha}
 G_\alpha(K)\equiv\frac{1}{K^2+\alpha^2}\,,
\eeq
and where $\smash{{\cal P}_{\mu\nu}^\perp(K)\equiv\delta_{\mu\nu}-K_\mu K_\nu/K^2}$. Note the identities $G^\kappa(K)=G^{-\kappa}(-K)$ and $G^\kappa_{\mu\nu}(K)=G^{-\kappa}_{\mu\nu}(-K)$, which follow from \eqn{eq:idmom} and which simply reflect the fact that the choice of orientation of the momentum/charge flow in the diagrams of \Fig{fig:propags} is arbitrary.

\begin{figure}[t!]  
\begin{center}
\epsfig{file=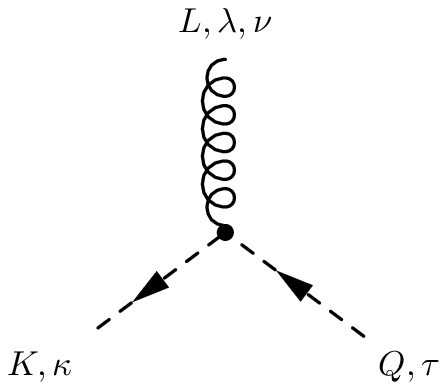,width=3.2cm}\qquad\epsfig{file=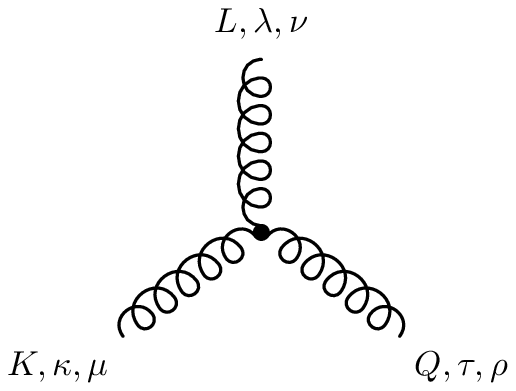,width=3.7cm}\\
 \caption{Diagrammatic representation of the cubic derivative vertices. The momenta and color charges are either all outgoing or all incoming.}\label{fig:vertex3}
\end{center}
\end{figure}

\begin{figure}[t!]  
\begin{center}
\epsfig{file=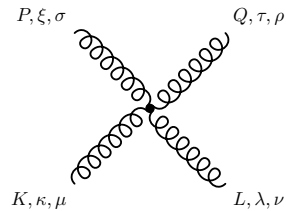,width=3.7cm}\\
 \caption{Diagrammatic representation of the four-gluon vertex. All momenta and color charges are either all outgoing or all incoming.}\label{fig:vertex4}
\end{center}
\end{figure}

The cubic (derivative) vertices are represented on \Fig{fig:vertex3}, with the convention that all momenta and color charges are outgoing. The ghost-antighost-gluon vertex is given by
\beq
\label{eq:v31}
 g\varepsilon^{\kappa\lambda\tau}K_\nu^\kappa,
\eeq
and the three-gluon vertex reads
\begin{align}
  \label{eq:v32}
\frac{g}{6}\varepsilon^{\kappa\lambda\tau}\Big[ \delta_{\mu\rho}\left(K^\kappa_\nu -Q^\tau_\nu \right)\!+\!\delta_{\nu\mu}\left(L^\lambda_\rho-K^\kappa_\rho\right)\!+\!\delta_{\rho\nu}\left(Q^\tau_\mu-L^\lambda_\mu \right)\!\Big],
\end{align}
where the various momenta, color charges, and spacetime indices are organized as in \Fig{fig:vertex3}. The structure constant $\varepsilon^{\kappa\lambda\tau}$ guarantees that the charge is conserved at the vertex: $\kappa+\lambda+\tau=0$. 
Finally, the four-gluon vertex is
\begin{align}
 \frac{g^2}{24}\sum_\omega\Big[\,&\varepsilon^{\kappa\lambda\omega}\varepsilon^{-\omega\tau\xi}(\delta_{\mu\rho}\delta_{\nu\sigma}-\delta_{\mu\sigma}\delta_{\nu\rho})\nonumber\\
  +&\varepsilon^{\kappa\tau\omega}\varepsilon^{-\omega\lambda\xi}(\delta_{\mu\nu}\delta_{\rho\sigma}-\delta_{\mu\sigma}\delta_{\nu\rho})\nonumber\\
  \label{eq:v4}
  +&\varepsilon^{\kappa\xi\omega}\varepsilon^{-\omega\tau\lambda}(\delta_{\mu\rho}\delta_{\nu\sigma}-\delta_{\mu\nu}\delta_{\rho\sigma})\,\Big],
\end{align}
where color charge and spacetime indices are organized as in \Fig{fig:vertex4}. Again, the combinations of the group structure constants guarantee that the color charge is conserved: $\kappa+\lambda+\tau+\xi=0$. 

Written in the basis \eqn{eq:basis}, the Feynman rules \eqn{eq:ghpropag}--\eqn{eq:v4} in the presence of the background field are very similar to the standard ones, usually written in the Cartesian color basis. The essential difference stems in the different structure constants, which can be traced back to the commutation relations \eqn{eq:basis}, and the role of the background field is simply to replace all momenta by shifted ones according to the corresponding color charges.\footnote{ The fact that the background field only appears explicitly through shifted momenta guarantees that the zero temperature contribution $V_{\rm vac}$ in \Eqn{eq:poteff} is independent of $r$. Indeed, the corresponding closed diagrams can be written in terms of continuous $d$-dimensional momentum integrals, and the various shifts in momentum can be absorbed in simple changes of variables (the conservation of the shifts is essential here). This is not possible in the finite temperature contributions, which involve a discrete sum over Matsubara frequencies.} As we shall discuss in a future work, these remarks generalize to any group SU($N$)---in fact to any compact Lie group with a semisimple Lie algebra. For $N=2$, one has the identities
\beq
\label{eq:identity1}
 \varepsilon^{(-\kappa)(-\lambda)(-\tau)}=-\varepsilon^{\kappa\lambda\tau}
\eeq
and 
\beq
\label{eq:identity2}
 \sum_\omega\varepsilon^{\kappa\lambda\omega}\varepsilon^{-\omega\tau\xi}=\delta^{\kappa,-\xi}\delta^{\lambda,-\tau}-\delta^{\kappa,-\tau}\delta^{\lambda,-\xi}.
\eeq
Using \eqn{eq:idmom} and \eqn{eq:identity1}, one checks that the vertices \eqn{eq:v31}--\eqn{eq:v4} are unchanged if all momenta and color charges are incoming.

%%%%%   The background field potential at two-loop order
\section{The background field potential at ${\cal O}(g^2)$}\label{sec:potential}

\begin{figure}[t!]  
\begin{center}
\epsfig{file=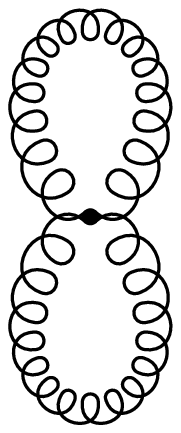,width=1.2cm}\quad\epsfig{file=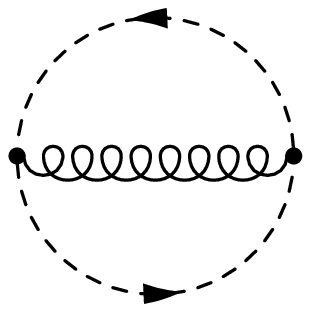,width=2.5cm}\quad\epsfig{file=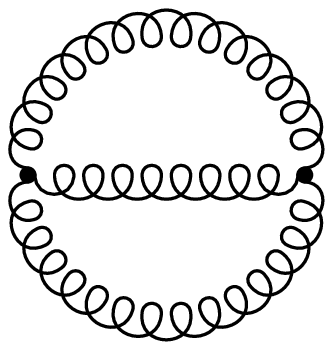,width=2.4cm}
 \caption{Two-loop diagrams contributing to the background field potential.}\label{fig:2loop}
\end{center}
\end{figure}

We consider the loop expansion of the background field potential, which corresponds to a perturbative expansion in powers of the renormalized coupling $g$ with $g\bar A_0\sim{\cal O}(1)$. We write the corresponding series as
\beq
\label{eq:pertexp}
 V(T,r)=\sum_{n\ge0}V^{(n)}(T,r)\,,
\eeq
with $\smash{V^{(n)}\sim{\cal O}(g^{2n-2})}$ the $n$-loop order contribution.  The classical action \eqn{eq_gf} evaluated at $c=\bar c=0$, $a_\mu=0$, and $\bar A_\mu(x)=\bar A_0\delta_{\mu0}$ vanishes identically since $\bar F_{\mu\nu}=0$. The tree-level contribution to the term $V_{\rm vac}$ in \eqn{eq:poteff} is thus trivially independent of the background field and the tree-level potential is trivial
\beq
 V^{(0)}(T,r)=0\,.
\eeq
The one-loop contribution has been obtained in \cite{Reinosa:2014ooa} and the relevant two-loop diagrams are shown in Fig.~\ref{fig:2loop}.

%%%   The one-loop contribution
\subsection{The one-loop contribution}

We briefly recall the result of \cite{Reinosa:2014ooa}. Introducing the function
\begin{align}\label{eq:Weiss0}
 {\cal F}_m(T,r)=\frac{T}{\pi^2}&\int_0^\infty \!\!dq\,q^2\Big\{\ln\left(1-e^{-\beta\varepsilon_q}\right)\nn
 &+\ln\left(1+e^{-2\beta\varepsilon_q}-2e^{-\beta\varepsilon_q}\cos r\right)\Big\},
\end{align}
which is such that, for $r\in[0,2\pi]$,
\beq\label{eq:Weiss}
{\cal F}_0(T,r)=\frac{T^4}{6}\left[\frac{(r-\pi)^4}{2\pi^2}-(r-\pi)^2+\frac{\pi^2}{10}\right],\nonumber\\
\eeq
the one-loop background field potential can be written as
\beq
\label{eq:onelooppot}
 V^{(1)}(T,r)=\frac{3}{2}{\cal F}_m(T,r)-\frac{1}{2}{\cal F}_0(T,r),
\eeq
where the first term on the right-hand side is the contribution from the massive gluons and the second one is due to the massless ghosts.\footnote{More precisely, this contribution arises from a partial cancellation between the ghost contribution and the $ a_\parallel- h$ sector, where $a_\parallel\propto \bar D_\mu a_\mu$ is the longitudinal component (with respect to the shifted momentum) of the gluon field. This sector only plays a role in the one-loop contribution to the background field potential.} It reduces to the well-known Weiss potential \cite{Weiss:1980rj,Gross:1980br} in the high temperature limit :
\beq\label{eq:one}
 V^{(1)}_{T\gg m}(T,r)\approx{\cal F}_0(T,r),
\eeq
whereas it gives a confining, inverted Weiss potential at low temperature, where the contribution from massive modes is suppressed \cite{Braun:2007bx,Fister:2013bh}:
\beq\label{eq:two}
 V^{(1)}_{T\ll m}(T,r)\approx-\frac{1}{2}{\cal F}_0(T,r).
\eeq

%%%   Two-loop diagrams
\subsection{Two-loop diagrams}
\label{sec:touloupe}

Let us start with the ghost-gluon sunset diagram  (second diagram of Fig.~\ref{fig:2loop}). A straightforward application of the Feynman rules derived above yields
\begin{align}
\label{eq:44}
V^{(2)}_{\rm 1g2gh} =\sum_{\kappa,\lambda,\tau}{\cal C}_{\kappa\lambda\tau}\Big[-&\frac{g^2}{2}\int_{Q,K} Q_\mu^\kappa {\cal P}^\perp_{\mu\nu}(L_\tau) K_\nu^\lambda\nn
&\times G_0(Q_\kappa)G_0(K_\lambda)G_m(L_\tau)\Big],
\end{align}
with $Q+K+L=0$ and $\kappa+\lambda+\tau=0$. We employ the general notations $Q_\mu\equiv(\omega_n,\bf q\,)$, with $\omega_n=2\pi nT$, and 
\beq
 \int_Q\equiv \mu^{2\epsilon}\,T\sum_{n\in\mathds{Z}}\int\frac{ d^{d-1}q}{(2\pi)^{d-1}} \equiv \mu^{2\epsilon}\,T\sum_{n\in\mathds{Z}}\int_{\bf q},
\eeq 
with $\mu$ an arbitrary mass scale and we recall that $d=4-2\epsilon$. We have also introduced the completely symmetric tensor 
\beq
\label{eq:tensor}
 {\cal C}_{\kappa\lambda\tau}=\left(\varepsilon_{\kappa\lambda\tau}\right)^2,
\eeq 
whose components vanish unless $\kappa+\lambda+\tau=0$ and are equal to one otherwise. It follows from Eqs.~\eqn{eq:identity1} and \eqn{eq:identity2} that
\beq
\label{eq:contraint3}
\sum_{\tau}{\cal C}_{\kappa\lambda\tau}=1-\delta_{\kappa\lambda},
\eeq
which imply ($N=2$)
\beq
\label{eq:contraint2}
\sum_{\lambda\tau}{\cal C}_{\kappa\lambda\tau}=N\quad{\rm and}\quad
\sum_{\kappa\lambda\tau}{\cal C}_{\kappa\lambda\tau}=N(N^2-1)\,.
\eeq

For a vanishing background field, the term within brackets in the right-hand side of \Eqn{eq:44} is nothing but the expression of the ghost-gluon sunset in the massive extension of the Landau gauge, up to the corresponding color factor $N(N^2-1)=6$. The latter is recovered using \eqn{eq:contraint2}. The expression \eqn{eq:44} illustrates that the two-loop perturbative diagram at nonvanishing background can be obtained from the corresponding one at $\bar A=0$ as follows: First, one writes the momentum integrals in the massive Landau gauge\footnote{There are various ways of writing the diagram in the absence of background. Here, one should only consider those expressions which follow from a direct application of the Feynman rules, without making use of symmetry properties such as $G(Q)=G(-Q)$, which may not be valid at nonvanishing background field. In contrast, one can use any manipulation which exploits the conservation of the momenta at the vertices, since this property is obeyed by the shifted momenta as well; see below.} in terms of three momenta $Q$, $K$, and $L$, up to the color factor $N(N^2-1)$; then, one shifts the momenta to $Q_\kappa$, $K_\lambda$, and $L_\tau$ and averages with the weight ${\cal C}_{\kappa\lambda\tau}$. This property can be anticipated from the Feynman rules described above and actually generalizes to any closed diagram, with an appropriate weight factor. The conservation of color charge at the interaction vertices implies the conservation of the shifted momenta under the loop integrals, e.g., $Q_\kappa+K_\lambda+L_\tau=0$ in the two-loop case above. This leads to important simplifications of the calculations in the Landau-DeWitt gauge, as detailed in the Appendixes. 
Let us finally mention that the expression \eqn{eq:44} and the above remarks actually generalize to SU($N$), where the tensor \eqn{eq:tensor} involves the appropriate structure constant. If \Eqn{eq:contraint3} is specific to the case $N=2$, the properties \eqn{eq:contraint2} are true for arbitrary $N$.

For later use, we rewrite \Eqn{eq:44} in a more compact form. Using the conservation of shifted momenta under the sum in \eqn{eq:44}, we have
\beq
 -Q_\mu^\kappa {\cal P}_{\mu\nu}^\perp(L_\tau)K_\nu^\lambda=\frac{Q^2_\kappa K^2_\lambda-(Q_\mu^\kappa K_\mu^\lambda)^2}{L_\tau^2}.
\eeq
Furthermore, writing
\beq
\label{eq:themegatrick}
 \frac{G_m(L)}{L^2}=\frac{1}{m^2}\left[G_0(L)-G_m(L)\right],
\eeq
we obtain
\beq
{ V^{(2)}_{\rm 1g2gh}=\frac{g^2}{2m^2}\sum_{\kappa,\lambda,\tau}{\cal C}_{\kappa\lambda\tau}\Big[I^{\kappa\lambda\tau}_{000}-I^{\kappa\lambda\tau}_{00m}\Big],}
\eeq
in terms of the two-loop integral
\beq
\label{eq:integral}
{ I^{\kappa\lambda\tau}_{\alpha\beta\gamma}\equiv \int_{Q,K} \!\!\big[Q^2_\kappa K_\lambda^2-(Q_\mu^\kappa K_\mu^\lambda)^2\big]G_\alpha(Q_\kappa)G_\beta(K_\lambda)G_\gamma(L_\tau),}
\eeq
which is needed only for the case $\kappa+\lambda+\tau=0$. In Appendix~\ref{appsec:reduc}, we rewrite these integrals in terms of simpler scalarlike loop integrals; see \Eqn{eq:scalarsunset} below. Here, we simply notice for later use that $I^{\kappa\lambda\tau}_{\alpha\beta\gamma}$ is invariant under the simultaneous permutation of the upper and lower indices, when the corresponding shifts of momenta add up to zero:
\beq
\label{eq:symint}
 I^{\kappa\lambda\tau}_{\alpha\beta\gamma}=I^{\kappa\tau\lambda}_{\alpha\gamma\beta}=\ldots \quad{\rm for}\quad\kappa+\lambda+\tau=0.
\eeq

 The two diagrams with purely gluonic loops can be treated in a similar way. We give more details in Appendix~ \ref{appsec:gluloupe} and simply state the results here. The double tadpole diagram (first diagram of Fig.~\ref{fig:2loop}) yields
\beq\label{eq:eight}
{ V^{(2)}_{\rm 2g}=\frac{g^2}{4}\sum_{\kappa,\lambda,\tau}{\cal C}_{\kappa\lambda\tau}\left[(d^2-3d+3)J_m^\kappa J_m^\lambda-J^\kappa_{\mu\nu}J^\lambda_{\mu\nu}\right],}
\eeq
where we have defined the tadpole integrals
\beq
\label{eq:tadpole}
 J_{\alpha}^\kappa\equiv \int_Q G_\alpha(Q_\kappa)
\eeq
and 
\beq
\label{eq:tadpu}
 J_{\mu\nu}^\kappa\equiv \int_Q \frac{Q^\kappa_\mu Q^\kappa_\nu}{Q^2_\kappa}G_m(Q_\kappa).
\eeq
As for the gluon sunset diagram (third diagram of Fig.~\ref{fig:2loop}), we obtain
\begin{align}
 V^{(2)}_{\rm 3g} &= -\frac{g^2}{4} \sum_{\kappa,\lambda,\tau}{\cal C}_{\kappa\lambda\tau}\Bigg\{J_m^\kappa J_m^\lambda-J^\kappa_{\mu\nu}J^\lambda_{\mu\nu}\nn
\label{eq:sunset} 
&-\frac{4}{m^2}\left[\left(d-\frac{5}{4}\right)I^{\kappa\lambda\tau}_{mmm}-(d-1)I^{\kappa\lambda\tau}_{mm0}+\frac{1}{4}I^{\kappa\lambda\tau}_{m00}\right]\Bigg\},\nonumber\\
\end{align}
where (\ref{eq:symint}) has been used.

%%%   Counterterm contribution
\subsection{Counterterm contribution}

The expressions derived in the previous section contain ultraviolet (UV) divergences. These are canceled by the counterterms from the original action, as we show explicitly in Appendix~\ref{appsec:final}. Here, we compute the relevant counterterm contributions to the background field potential. 

Writing the renormalization factors as $Z_i=1+\delta Z_i$, $i=m^2,g,a,c$, the counterterm action reads
\begin{align}
 \delta S=\int_x\tr \Big\{&\!-\delta Z_a\,a_\mu(\delta_{\mu\nu}\bar D^2-\bar D_\mu\bar D_\nu)a_\nu+\delta m^2a_\mu a_\mu\nn
 &\!-2\delta Z_c\,\bar c\,\bar D^2 c+2ih\bar D_\mu a_\mu\Big\}+\ldots
\end{align}
where $\delta m^2=m^2(\delta Z_{m^2}+\delta Z_a)$ and the dots denote terms involving the coupling counterterm $\delta g$, which are not needed in the present work.

Using \eqn{eq:barderivative}, we obtain, for the ghost counterterm loop,
\beq
V^{(2)}_{\delta Z_c} = -\delta Z_c\sum_\kappa\int_Q  Q^2_\kappa G_0(Q_\kappa)=0\,,
\eeq
where the last integral vanishes in dimensional regularization. The gluon counterterm loop reads
\begin{align}
V^{(2)}_{\delta Z_A,\delta m^2} & =  \frac{1}{2}\sum_\kappa\int_Q\Big(\delta Z_aQ^2_\kappa+\delta m^2\Big) G^{\kappa}_{\mu\mu}(Q)\nonumber\\
& =  \frac{d-1}{2}\sum_\kappa \int_Q\Big(\delta Z_aQ^2_\kappa+\delta m^2\Big) G_m(Q_\kappa)\,.
\end{align}
Then, up to an integral which vanishes in dimensional regularization, the total counterterm contribution can be written in terms of the tadpole integral \eqn{eq:tadpole} as
\beq
 V^{(2)}_{\rm ct}=\frac{d-1}{2}m^2\delta Z_{m^2}\sum_\kappa J_m^\kappa\,.
\eeq

%%%   Total two-loop contribution
\subsection{The two-loop contribution}\label{sec:total}

Adding together the various two-loop pieces, we obtain [notice that the contributions from the integral \eqn{eq:tadpu} cancel]
\begin{align}
&V^{(2)}  =  \frac{d-1}{2}m^2\delta Z_{m^2}\sum_\kappa J_m^\kappa\nonumber\\
&\qquad +  g^2\sum_{\kappa,\lambda,\tau}{\cal C}_{\kappa\lambda\tau}\left\{\frac{(d-1)(d-2)}{4}J_m^\kappa J_m^\lambda+\frac{1}{2m^2}I^{\kappa\lambda\tau}_{000}\right.\nonumber\\
\label{eq:touloupewithI}
& \qquad +\left.\frac{1}{m^2}\left[\left(d-\frac{5}{4}\right)I^{\kappa\lambda\tau}_{mmm}-(d-1)I^{\kappa\lambda\tau}_{mm0}-\frac{1}{4}I^{\kappa\lambda\tau}_{m00}\right]\right\}.
\end{align}
For a vanishing background field, \Eqn{eq:touloupewithI} is nothing but the two-loop free energy density computed in the massive extension of the Landau gauge. As detailed in Appendix~\ref{appsec:reduc}, the integral \eqn{eq:integral} can be expressed in terms of the tadpole integrals \eqn{eq:tadpole} and
\beq
\label{eq:tadpoleq0}
\tilde J_{\alpha}^\kappa\equiv \int_Q Q_\kappa^0 G_\alpha(Q_\kappa),
\eeq
and of the scalar sunset
\beq
\label{eq:scalarsunset}
S^{\kappa\lambda\tau}_{\alpha\beta\gamma}  \equiv  \int_{Q,K} G_\alpha(Q_\kappa)G_\beta(K_\lambda)G_\gamma(L_\tau)\,.
\eeq
Using Eqs. \eqn{appeq:relation1}--\eqn{appeq:relation4}, our final expression for the two-loop contribution to the background field potential is
\begin{widetext}
\begin{align}
{ V^{(2)}} & = { \frac{d-1}{2}m^2\delta Z_{m^2}\sum_\kappa J_m^\kappa} +g^2{ \sum_{\kappa,\lambda,\tau}{\cal C}_{\kappa\lambda\tau}\Bigg\{\frac{1}{4}\left(d^2-4d+\frac{15}{4}\right)J_m^\kappa J_m^\lambda+\frac{1}{8}J_0^\kappa J_m^\lambda-\frac{1}{16}J_0^\kappa J_0^\lambda}\nonumber\\
\label{eq:fff}
& -\frac{1}{m^2} \left(d-\frac{11}{8}\right)\tilde J_m^\kappa \tilde J_m^\lambda+\frac{1}{m^2} \left(d-\frac{3}{4}\right)\tilde J_0^\kappa \tilde J_m^\lambda-\frac{5}{8m^2}\tilde J_0^\kappa \tilde J_0^\lambda+ \frac{3m^2}{4}\left(d-\frac{5}{4}\right) S_{mmm}^{\kappa\lambda\tau}+\frac{m^2}{16} S^{\kappa\lambda\tau}_{m00}\Bigg\}\,.
\end{align}
\end{widetext}
The calculation of the various Matsubara sums and momentum integrals involved in this expression is detailed in Appendix~ \ref{appsec:matsubara}. Moreover, we show in Appendix~\ref{appsec:final} that it is UV finite (up to an overall, temperature- and background-independent divergence) once the counterterm $\delta Z_{m^2}$ has been fixed from the renormalization conditions \eqn{eq:ren_cond}. We also reduce this expression to a sum of one- and two-dimensional (radial) momentum integrals involving thermal, Bose-Einstein distribution functions in the presence of the background field. Our final result for the thermal part (the only one that depends on the background) of the two-loop contribution to the background field potential is given in \Eqn{appeq:final}.

%%%%%   The Polyakov loop at one-loop order
\section{The Polyakov loop at ${\cal O}(g^2)$}\label{sec:polyakov_loop}

We similarly expand the Polyakov loop \eqn{eq:ploop} as
\beq
 \ell(T)=\sum_{n\ge0}\ell^{(n)}(T),
\eeq
with the $n$-loop contribution $\ell^{(n)}\sim{\cal O}(g^{2n})$. This is obtained by expanding the path-ordered exponential in \Eqn{eq:ploop} in powers of the coupling with $g\bar A_0\sim{\cal O}(1)$. The tree-level and one-loop contributions are evaluated in Appendix~\ref{appsec:PL} for arbitrary $N$ and for fields in an arbitrary representation of the gauge group. For the fundamental representation of SU(2), we obtain\footnote{This is the renormalized Polyakov loop in dimensional regularization.}
\beq
\label{eq:oth}
 \ell^{(0)}(T)=\cos\left(\frac{r_{\rm min}(T)}{2}\right)
\eeq
and
\beq
\frac{\ell^{(1)}(T)}{\ell^{(0)}(T)}=g^2\beta m\left[\frac{3}{32\pi}+\frac{a(T,r_{\rm min}(T))}{4\pi^2}\sin^2\left(\frac{r_{\rm min}(T)}{2}\right)\right]
\eeq
where $r_{\rm min}(T)$ is the absolute minimum of the two-loop background field potential and
\beq
a(T,r)=-\int_0^\infty\frac{k^2dk}{m^3}\left\{\frac{k^2/\varepsilon^2_k}{\cosh(\beta\varepsilon_k)-\cos r}-(m\to 0)\right\}.
\eeq
It is easily checked that $a(T,r)\geq 0$; hence
\beq
 \ell_{\rm 1loop}(T)=0\, \Leftrightarrow\,\ell^{(0)}(T)=0 \,\Leftrightarrow \,r_{\rm min}(T)=\pi\,({\rm mod}\, 2\pi).
\eeq
This confirms, at this order, that the background field $r$ itself is a good order parameter for static quark confinement \cite{Braun:2007bx,Braun:2010cy,Fister:2013bh}.

%%%%%   Results   
\section{Results}\label{sec:results}

\begin{figure}[t!]  
\begin{center}
\epsfig{file=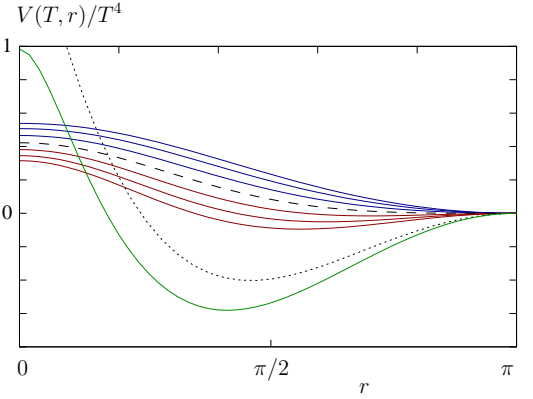,width=7cm}
 \caption{Rescaled two-loop background field potential $V(T,r)/T^4$ for various temperatures, below (blue) and above (red) the critical temperature (dashed black). The green curve corresponds to a higher temperature and shows the approach to the asymptotic high temperature limit $v_\infty(r)$ (dotted line); see \Eqn{eq:asymptotic}. All curves have been shifted by their respective values at $r=\pi$ for clarity.}\label{fig:pot}
\end{center}
\end{figure}

As already mentioned, we shall use the values of $m$ and $g$ inferred from lattice calculations of ghost and gluon propagators in the Landau gauge at vanishing temperature. For the present order of approximation, we fit these data to the one-loop expressions obtained in the massive Landau gauge. The best fitting values are $m\simeq 680$~MeV and $g\simeq 7.5$ with $\mu=1$~GeV. { This sets the scale of the present calculation.} We shall compare our findings to our earlier leading-order results \cite{Reinosa:2014ooa} where the mass parameter was obtained by fitting the data to the tree-level propagators with the best fitting value $m\simeq 710$~MeV. There is of course some error related to the determination of the parameters. We have checked that the results to be presented below do not change qualitatively as we change the parameters within the error bars. Let us finally mention that we shall not consider the possible temperature dependence of the parameters.\footnote{This has been studied in the case of the Landau gauge by fitting one-loop correlators against lattice data at finite temperature \cite{Reinosa:2013twa}. It was found that the parameter $m$ is essentially independent of the temperature whereas the coupling $g$ shows a $30$\% decrease between $T=0$ and $T_c$. Note, however, that these numbers were obtained in the Landau gauge and are not directly applicable here. Still, for completeness, we have checked that a $30$\% change in the coupling results in a $10$\% change in the critical temperature.} This and the implementation of renormalization-group improvement are deferred to a future work.

%%%   Background potential
\subsection{Background field potential}

\begin{figure}[t!]  
\begin{center}
\epsfig{file=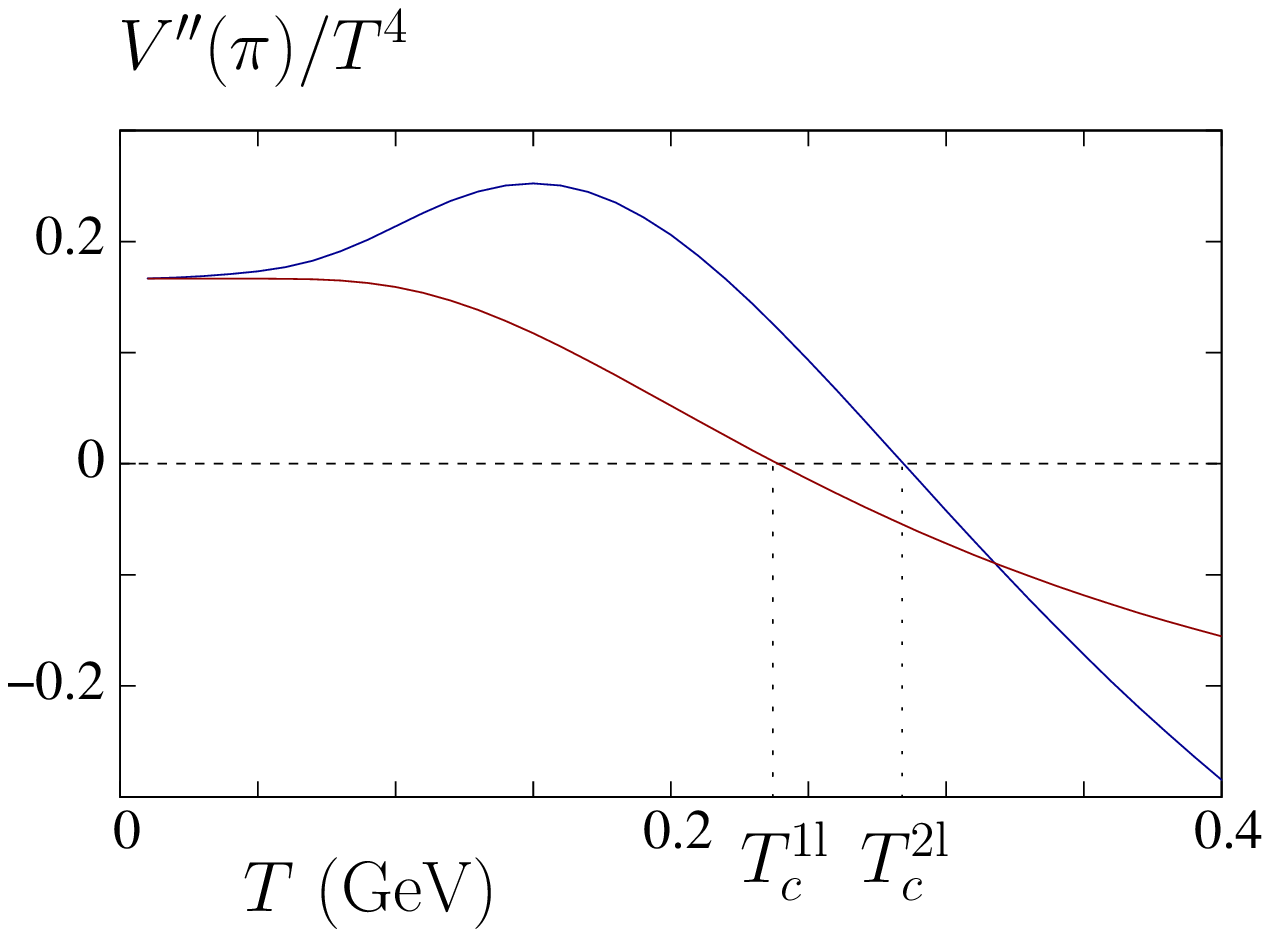,width=4.35cm}\,\,\,\epsfig{file=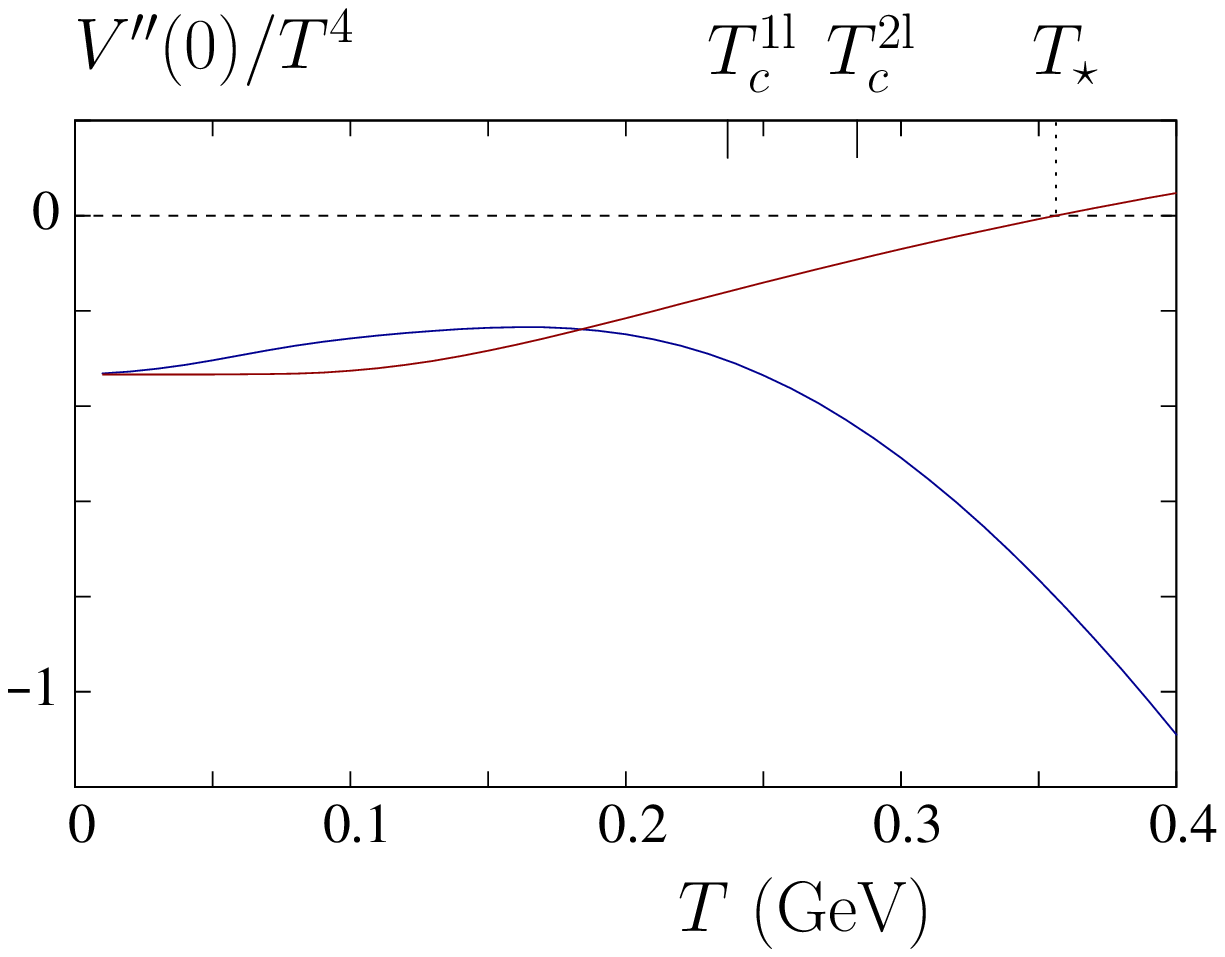,width=4.1cm}
 \caption{Rescaled curvature of the one-loop (red) and two-loop (blue) background potentials at $r=\pi$ (left) and $r=0$ (right) as functions of the temperature. The curvatures at $r=\pi$ vanish at the corresponding critical temperatures, denoted here by $T_c^{1l}$ and $T_c^{2l}$, respectively. The one-loop curvature at $r=0$ vanishes at a temperature $T_\star$, above which the minimum of the one-loop potential is at $r=0$. This does not happen at two loops.}\label{fig:curv}
\end{center}
\end{figure}

In Fig.~\ref{fig:pot}, we show the rescaled two-loop background field potential $V(T,r)/T^4$ in the range $[0,\pi]$ for various values of the temperature. As the temperature is increased, there is clearly a transition from a confining phase, where the minimum of the potential lies at its confining value $r=\pi$, to a deconfined phase where the location of the minimum departs from $r=\pi$. The transition is second order and the corresponding critical temperature is obtained by requiring the vanishing of the curvature of the potential at the confining point. The rescaled curvature $\partial_r^2V(T,r)/T^4$ at $r=\pi$ is plotted as a function of $T$ in the left panel of Fig.~\ref{fig:curv} at one- and two-loop orders. The two-loop correction leads to a larger critical temperature, $T_{\rm c}^{\rm 2loop}\simeq 284\,{\rm MeV}$, as compared to our previous one-loop result  \cite{Reinosa:2014ooa}, $T_{\rm c}^{\rm 1loop}\simeq 237\,{\rm MeV}$. A typical lattice result is \cite{Lucini:2012gg} $T_c^{\rm latt}=295$~MeV. Although such comparison must  be taken with care due to the issue of properly setting the scale, this shows that the two-loop corrections indeed improve the one-loop result. 

It is also interesting to compare with other continuum approaches. The FRG and DSE/2PI calculations of Ref.~\cite{Fister:2013bh} give $T_c^{\rm FRG}=230\,{\rm MeV}$ and $T_c^{\rm DSE/2PI}=235\,{\rm MeV}$, respectively, which lie in the same ballpark as our one-loop result. The improved value of the critical temperature obtained above suggests that the present two-loop calculation efficiently captures some of the effects which have been discarded in those calculations. For instance, although the fully resummed propagators are included, some explicit two-loop contributions to the DSE for the background field potential have been neglected. As for the FRG calculation, the authors of Ref.~\cite{Fister:2013bh} mention that their result is modified to $T_c^{\rm FRG}=300\,{\rm MeV}$ when some backreaction effects---neglected in their main study---are included.

\begin{figure}[t!]  
\begin{center}
\epsfig{file=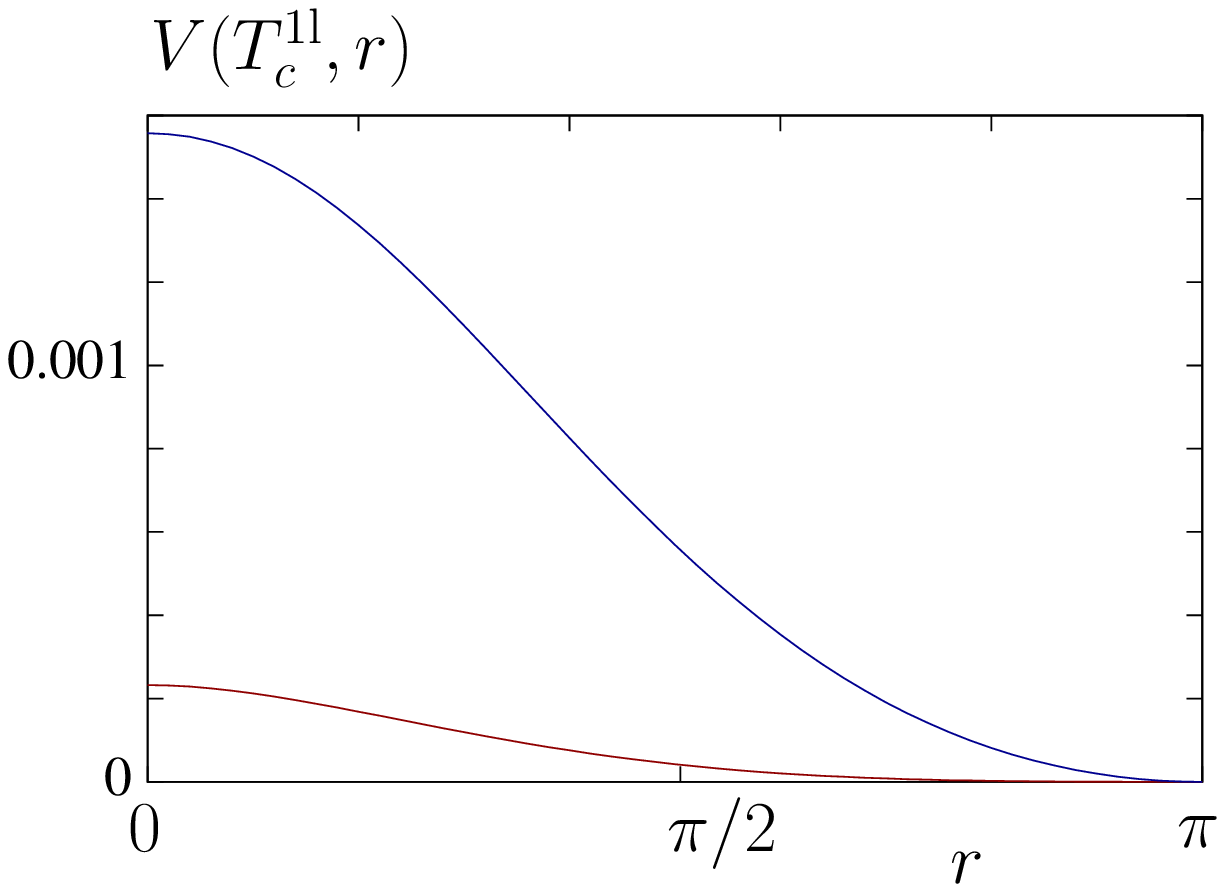,width=4.3cm}\epsfig{file=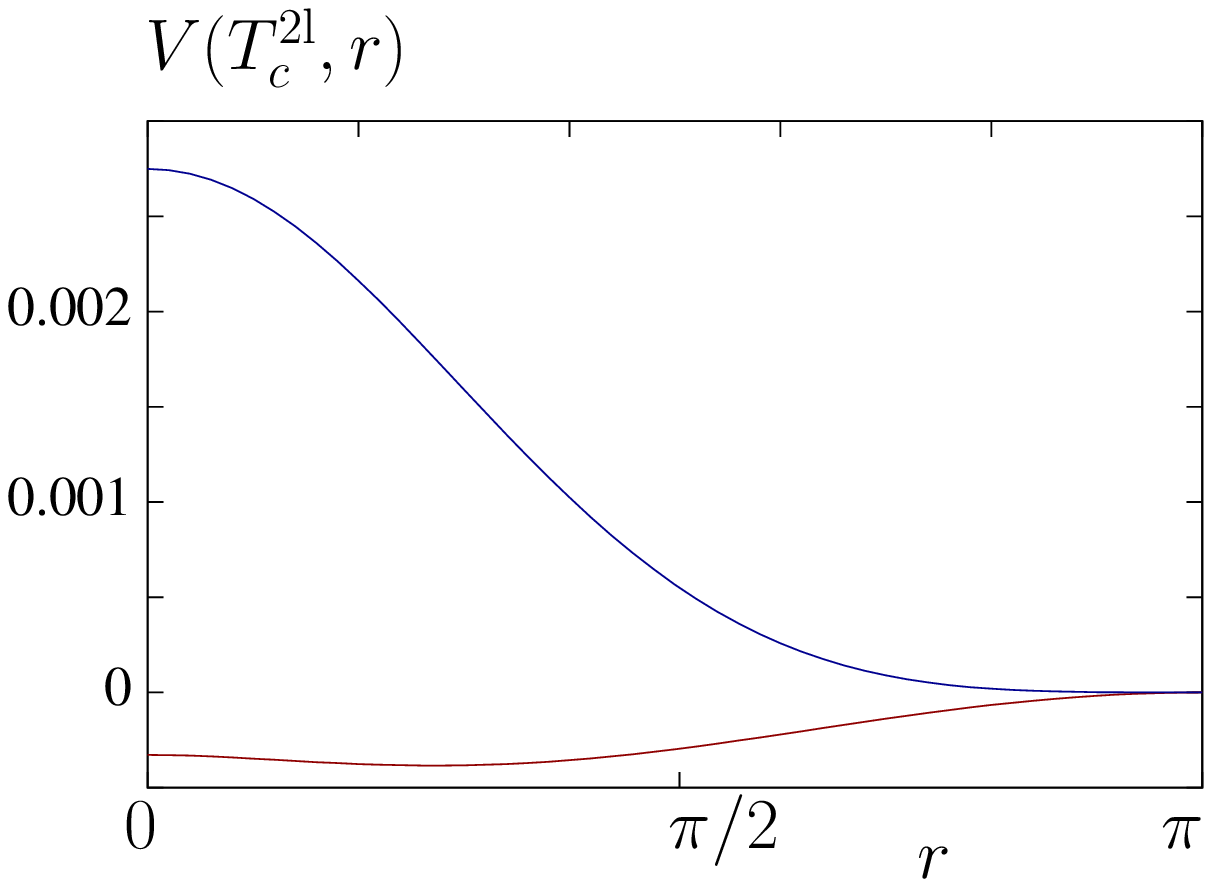,width=4.3cm}
 \caption{One-loop (red) and two-loop (blue) background field potentials as functions of $r$ at $T=T_{\rm c}^{\rm 1loop}$ (left) and $T=T_{\rm c}^{\rm 2loop}$ (right). The curves have been shifted by their respective values at $r=\pi$ for clarity.}\label{fig:pot_comp}
\end{center}
\end{figure}

For completeness, we compare in Fig.~\ref{fig:pot_comp} the one- and two-loop potentials at their respective critical temperatures. It is also instructive to plot the (rescaled) curvature of the potential at the origin as a function of $T$; see the right panel of Fig.~\ref{fig:curv}. For the values of parameters studied here, we observe that, at one-loop order, there exists a temperature $T_\star\simeq 1.5 T^{\rm 1loop}_{\rm c}$ above which the minimum of the potential is exactly located at $r=0$. This does not seem to be the case at two-loop order where the curvature of the potential at the origin remains negative. 

To have a better analytical control on our results, we have considered the formal limit $T\to\infty$.\footnote{Here, this limit is used to control our semianalytical results and to illustrate the difference between the one-loop and two-loop calculations at fixed coupling. It is by no means a physically relevant limit because important (high temperature) effects such as the running of the coupling, or the physics of hard thermal loops are not included in the present calculation. In particular, if the running of the coupling is taken into account in (\ref{eq:rinf}), one obtains $r_\infty\to0$ [see \Eqn{eq:rinf}] at asymptotically high temperatures { and one recovers the Weiss potential.}} We show in Appendix \ref{appsec:rinf} that the rescaled potential $v_\infty(r)=\lim_{T\to\infty} V(T,r)/T^4$ is a polynomial in the range $[0,\pi]$:
\bea
v_\infty(r) & = & \frac{\pi^2}{60}\left[5\left(\frac{r}{\pi}-1\right)^4-10\left(\frac{r}{\pi}-1\right)^2+1\right]\nonumber\\
\label{eq:asymptotic}
& + & \frac{g^2}{96}\left[7\left(\frac{r}{\pi}-1\right)^4-2\left(\frac{r}{\pi}-1\right)^2-1\right],
\eea
which minimum is located at 
\beq\label{eq:rinf}
r_\infty=\pi\left(1-\sqrt{\frac{8\pi^2+g^2}{8\pi^2+7g^2}}\right).
\eeq
We discuss the consequences of the different behavior between the one- and two-loop results below. 

%%%   Polyakov loop vs temperature
\subsection{Polyakov loop}

\begin{figure}[t!]  
\begin{center}
\epsfig{file=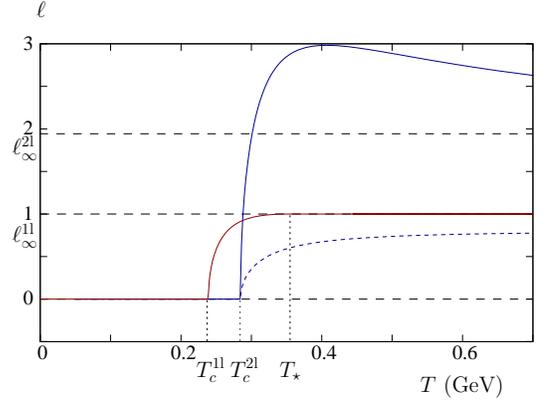,width=7cm}
 \caption{Temperature dependence of the Polyakov loop at leading (red) and next-to-leading (blue) orders. The horizontal dashed lines denote the corresponding asymptotic values at high temperature, denoted here by $\ell_\infty^{1l}$ and $\ell_\infty^{2l}$, respectively. The dashed blue curve shows the mean-field Polyakov loop \eqn{eq:oth} evaluated at $r=r_{\rm min}^{\rm 2loop}$.The respective one- and two-loop critical temperatures are indicated by vertical dashed lines, as well as the temperature $T_\star$ at which the leading-order Polyakov loop reaches its asymptotic value.}\label{fig:pl}
\end{center}
\end{figure}

In Fig.~\ref{fig:pl}, we compare the temperature dependence of the Polyakov loop at leading\footnote{The Polyakov loop at leading order is evaluated from \Eqn{eq:oth}, with $r_{\rm min}(T)$ the minimum of the leading-order potential \eqn{eq:onelooppot}.} and next-to-leading order. At leading order, it saturates to its upper bound $\ell^{(0)}_\infty=1$ at the temperature $T_\star$ discussed previously, above which $r=0$ remains the absolute minimum of the potential; see Fig.~\ref{fig:curv}. The Polyakov loop is singular at $T=T_\star$ because its second derivative is discontinuous. This has to do with the particular form of the Weiss potential, \Eqn{eq:Weiss}. For small positive values of $r$, the latter behaves as
\beq
{\cal F}_0(T,r)=T^4\left[-\frac{\pi^2}{15}+\frac{r^2}{3}-\frac{r^3}{3\pi}\dots\right]
\eeq
and is thus nonanalytic in $r^2$. This is to be contrasted with the massive version ${\cal F}_m(T,r)$ of the same function which shows a regular expansion in powers of $r^2$ when $m\neq 0$. In the vicinity of $r=0$ and for $T$ close to $T_\star$, the rescaled potential is of the form $a(T)+b(T) r^2+c r^3$, where $b(T)=b_-(T_\star-T)$ with $b_-<0$ and $c>0$. It follows, that for $T$ approaching $T_\star$ from below,  $r(T)\propto T_\star-T$ and $r(T)=0$ for $T>T_\star$. Thus, the first derivative of the background with respect to the temperature is discontinuous at $T=T_\star$. This singularity propagates to thermodynamic observables. For instance the third derivative of the free energy density with respect to the temperature is discontinuous. This is, however, a spurious discontinuity.

As already discussed above, at two-loop order the curvature of the potential at the origin remains negative. In this range, the Polyakov loop does not show any additional singularity, other than the one at $T^{\rm 2loop}_{\rm c}$. We also observe that, as compared to the one-loop result, the Polyakov loop at two-loop order overshoots its asymptotic ($T\to\infty$) value
\beq
\ell^{\rm 1loop}_\infty= \cos\left(\frac{r_\infty }{2}\right)+\frac{3g^2}{16\pi^2}(\pi-r_\infty)\sin\left(\frac{r_\infty }{2}\right),
\eeq
with $r_\infty$ given in (\ref{eq:rinf}), as computed in Appendix~\ref{appsec:PL}.

%%%   Pressure and entropy
\subsection{Pressure and entropy}

\begin{figure}[t!]  
\begin{center}
\epsfig{file=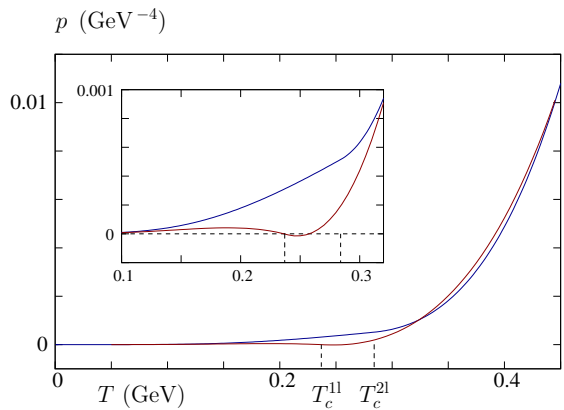,width=7.5cm}
 \caption{Thermodynamic pressure at one- (red) and two-loop (blue) orders, obtained from the minimum of the background field potential as a function of the temperature. The plot in the inset is a zoom on the low temperature region. The respective one- and two-loop critical temperatures are indicated by vertical dashed lines.}\label{fig:pressure}
\end{center}
\end{figure}

The thermodynamic pressure \eqn{eq:pressure} is shown in Fig.~\ref{fig:pressure} as a function of the temperature in the one- and two-loop approximations. In both cases, we observe that the pressure is increasing at small temperatures, indicating a positive entropy ($s=dp/dT$). This is a welcome result although it may be surprising at first sight because the ghosts dominate in this regime and one would naively think that they contribute negatively to the entropy. 

\begin{figure}[t!]  
\begin{center}
\epsfig{file=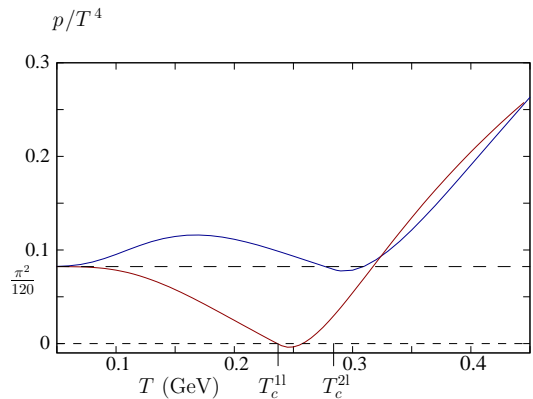,width=7.cm}
 \caption{Thermodynamic pressure rescaled by $T^4$, at one-loop (red) and two-loop (blue) orders, as a function of temperature.}\label{fig:povert4}
\end{center}
\end{figure}

The positivity of the entropy at low temperatures is ensured by the fact that, in the confining phase with $r=\pi$, those ghosts which feel the background effectively behave as physical fermions, giving a positive contribution to the entropy. To illustrate this point more precisely, we note that the low temperature behavior of the background field potential is dominated by the one-loop contribution, as discussed in Appendix~\ref{appsec:final}. This is directly visible in Fig.~\ref{fig:povert4}. At one-loop order, in the confined phase, the entropy contribution $\Delta_\kappa s$ of a ghost with charge $\kappa$ is
\beq
\frac{\Delta_\kappa s}{2T^3}=\int_{\bf q}\ln\Big(1+e^{-2q}-2e^{-q}\cos(\kappa\pi)\Big).
\eeq
So we have either a bosoniclike contribution ($d=4$)
\beq
\frac{\Delta_0 s}{4T^3}=\int_{\bf q}\ln\Big(1-e^{-q}\Big)=-\frac{\pi^2}{90}\,,\\
\eeq
with a standard negative (ghostlike) sign from the neutral modes, or fermioniclike contributions
\beq
\frac{\Delta_\pm s}{4T^3}=\int_{\bf q}\ln\Big(1+e^{-q}\Big)=\frac{7\pi^2}{720}
\eeq
with a positive sign for charged modes due to their coupling to the background.\footnote{A similar discussion can be done in terms of thermal distribution functions. Those modes which are not affected by the background are associated with bosonic distribution functions $n_{\varepsilon}$, whereas those modes which are affected by the background are associated with shifted distribution functions ${\rm Re}\,n_{\varepsilon-irT}$ which, up to a sign, become fermionic distribution functions ${\rm Re}\,n_{\varepsilon-i\pi T}=-f_\varepsilon$ when the background takes the confining value $r=\pi$.} The total ghost contribution,
\beq
\frac{\Delta_0 s+2\Delta_+ s}{4T^3}=\frac{\pi^2}{120},
\eeq
 is then positive, as announced.

As one increases the temperature, the one-loop result violates the positivity of the entropy, slightly before reaching $T_{\rm c}^{\rm 1loop}$, as can be clearly seen in the inset plot of Fig.~\ref{fig:pressure}, where the thermal pressure changes its monotony and even becomes slightly negative before $T_{\rm c}^{\rm 1loop}$. The reason for this behavior is again the change of effective statistics of the relevant degrees of freedom in the presence of the background. As the temperature is increased, the massive gluons start contributing to the pressure. However, in the confined phase, the charged gluons, which feel the presence of the background, behave like ``wrong'' fermions, contributing negatively to the entropy \cite{Sasaki:2012bi}. Remarkably this behavior is completely washed out by the two-loop correction and at two-loop order the entropy is positive (the pressure is monotonically increasing with the temperature) in the whole range of temperatures studied here; see Fig.~\ref{fig:pressure}.

Finally, let us comment on the $T^4$ behavior of the pressure at low temperature---see Fig.~\ref{fig:povert4}---which is at odds with the exponential suppression seen in lattice results~\cite{Engels:1981qx}. As discussed above, this originates from the fact that massless (ghost) modes directly contribute to the pressure, already at leading order. We mention that this is likely to be a general problem for continuum approaches, which are essentially based on using (resummed) propagators. For instance, in the Landau gauge, lattice results for the propagators \cite{Boucaud:2011ug,Maas:2011se} show that, if the gluon becomes effectively massive for infrared momenta, the ghost remains massless. This generically produces $T^4$ contributions in a leading-order-like---i.e., trace-log---expression for the thermodynamic pressure. The correct treatment of such unphysical massless degrees of freedom is a serious issue that needs to be further investigated.

%%%
\subsection{Comparison with the massive Landau gauge}

\begin{figure}[t!]  
\begin{center}
\epsfig{file=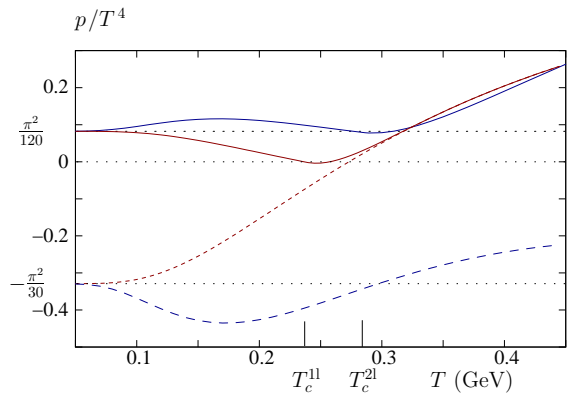,width=7.5cm}
 \caption{Comparison between one-loop (red) and two-loop (blue) results in the massive Landau-DeWitt gauge (plain) and in the massive Landau gauge (dashed). Both one-loop results coincide for $T>T_\star$ (see text) because in that case $r=0$. The critical temperatures indicated on the horizontal axis are those obtained from the Landau-DeWitt calculation.}\label{fig:Landau}
\end{center}
\end{figure}

It is interesting to compare our results for the pressure to those obtained in the massive Landau gauge (which corresponds to a vanishing background). The corresponding one- and two-loop pressure curves are shown in Fig.~\ref{fig:Landau} and compared to those discussed in the previous subsection. We observe again that the one-loop pressure contains $T^4$ contributions at low temperatures but this time those come with a negative prefactor which yields a negative entropy at low temperatures. The reason for this is simple: in the absence of background all ghosts contribute negatively to the entropy. In this case the two-loop correction is of no help because it does not contain any $T^4$ contribution at small temperature. In fact it seems that the two-loop term makes the problem even worse since the entropy remains negative in the range of temperature shown here. Furthermore, we observe in Fig.~\ref{fig:Landau} that the two-loop correction is smaller in the Landau-DeWitt case than in the Landau case, indicating a better convergence.

The above remarks illustrate the importance of taking into account the order parameter of the $Z_N$ transition in the description. Our two-loop results show that the value $r=0$---which would correspond to the Landau gauge---is never a physical point, i.e., an absolute minimum of the background field potential. In a sense, the perturbative expansion in the (massive) Landau gauge appears as an expansion around an unstable situation. It would take infinite resummations to correctly describe the physics near the stable physical point.We believe that this remark outranges the present framework and holds for other continuum approaches as well. 

%%%%%   Conclusions
\section{Conclusions}

To summarize, we have proposed in \cite{Reinosa:2014ooa} a perturbative approach to describe the static quark confinement-deconfinement transition in Yang-Mills theories, based on a modified (massive) gauge-fixed action in the Landau-DeWitt gauge. This describes well the phase structure of SU($N$) theories at leading order and gives qualitatively good results for the transition temperatures of the $N=2$ and $N=3$ cases in $d=4$. However, this leads to a $T^4$ behavior of the thermodynamic pressure at low temperatures and to negative entropy and negative pressure near the transition temperature. Also the leading-order calculation leads to a spurious singularity at a temperature $T_\star$ above which the physical background field vanishes.

With the present work, we wish to demonstrate the interest and feasibility of a next-to-leading-order calculation in this modified perturbative scheme. We investigate the SU($2$) theory at next-to-leading order, which shows that the correction to the critical temperature goes in the right direction and actually brings the estimated value close to the lattice result---although we stress again that such a comparison must be taken with care due to the issue of scale setting. The next-to-leading-order corrections also cure the negative entropy issue and remove the spurious singularity at $T=T_\star$. In fact, we find that the physical point always corresponds to a nonvanishing background field.  There remains the issue of unphysical $T^4$ contributions to the pressure from the massless (ghost) degrees of freedom, which needs to be investigated.

In a future work, we shall extend the present two-loop study to the SU($3$) theory and investigate the effect of renormalization-group improvement. Another interesting extension would be to include quark degrees of freedom at nonzero chemical potential. Finally, the present work points out the importance of taking into account a nonvanishing background field near the phase transition. We plan to extend our previous study \cite{Reinosa:2013twa} of the Landau gauge ghost and gluon correlators at finite temperature to the Landau-DeWitt gauge.

\section*{ACKNOWLEDGEMENTS}

We thank  J.~M. Pawlowski for interesting discussions. We acknowledge financial support from the PEDECIBA and ECOS programs. U.R., J.S., and M.T. thank the IFFI, and N.W.  the LPTMC for their hospitality.

\appendix

%%%%%   Gluon loops
\section{Gluon loops}
\label{appsec:gluloupe}

Here, we detail the derivation of Eqs. \eqn{eq:eight} and \eqn{eq:sunset} for the purely gluonic two-loop diagrams. A straightforward application of the Feynman rules of Sec.~\ref{sec:Frules} yields, for the double gluon tadpole diagram  (first diagram of Fig.~\ref{fig:2loop}),
\begin{align}
V^{(2)}_{\rm 2g}  =- \frac{g^2}{8}&\sum_{\kappa,\lambda,\omega}\int_{Q,K} G_{\mu\nu}(Q_\kappa) G_{\rho\sigma}(K_\lambda)\nn
\times \Big[\,\,&\varepsilon^{\kappa(-\kappa)\omega}\varepsilon^{(-\omega)\lambda(-\lambda)}(\delta_{\mu\rho}\delta_{\nu\sigma}-\delta_{\mu\sigma}\delta_{\nu\rho})\nonumber\\
+\,&\varepsilon^{\kappa\lambda\omega}\varepsilon^{(-\omega)(-\kappa)(-\lambda)}(\delta_{\mu\nu}\delta_{\rho\sigma}-\delta_{\mu\sigma}\delta_{\nu\rho})\nonumber\\
+\,&\varepsilon^{\kappa(-\lambda)\omega}\varepsilon^{(-\omega)\lambda(-\kappa)}(\delta_{\mu\rho}\delta_{\nu\sigma}-\delta_{\mu\nu}\delta_{\rho\sigma})\Big].
\end{align}
The contribution from the first line in the brackets vanishes and the two other lines contribute the same. Setting $\lambda\to-\lambda$ and $K\to-K$ in the third line and using \eqn{eq:idmom} and $G_{\rho\sigma}(-K_\lambda)=G_{\rho\sigma}(K_\lambda)$, we obtain
\begin{align}
\label{appeq:2gfirst}
V^{(2)}_{\rm 2g} & = \frac{g^2}{4}\sum_{\kappa,\lambda,\tau}{\cal C}_{\kappa\lambda\tau} \int_{Q,K}G_m(Q_\kappa)G_m(K_\lambda)\nn
&\times\Big[{\cal P}^\perp_{\mu\mu}(Q_\kappa){\cal P}^\perp_{\nu\nu}(K_\lambda)-{\cal P}^\perp_{\mu\nu}(Q_\kappa){\cal P}^\perp_{\mu\nu}(K_\lambda)\Big],
\end{align}
where we renamed $\omega\to\tau$ and we used the definition \eqn{eq:tensor}. This rewrites as \Eqn{eq:eight} using the integrals \eqn{eq:tadpole} and \eqn{eq:tadpu}. As mentioned in Sec. \ref{sec:touloupe} for the ghost-gluon sunset diagram, we observe that the expression \eqn{appeq:2gfirst} can be obtained from the corresponding diagram at vanishing background field by replacing the loop momenta by shifted ones and by averaging with the weight ${\cal C}_{\kappa\lambda\tau}$, where, here, the index $\tau$ is redundant since it is not associated to any internal momentum.

A similar writing applies to the gluon sunset diagram (third diagram of Fig.~\ref{fig:2loop}). Applying the Feynman rules of Sec.~\ref{sec:Frules}, we obtain, after simple manipulations, 
\begin{widetext}
\beq\label{eq:49}
V^{(2)}_{\rm 3g} = -g^2\sum_{\kappa,\lambda,\tau}{\cal C}_{\kappa\lambda\tau}\int_{Q,K} {\cal B}(Q_\kappa,K_\lambda,L_\tau) G_m(Q_\kappa)G_m(K_\lambda)G_m(L_\tau),
\eeq
where $Q+K+L=0$ and
\beq
 {\cal B}(Q,K,L)=\left[Q\cdot {\cal P}^\perp(L)\cdot Q\right]\tr\left[{\cal P}^\perp(K) {\cal P}^\perp (Q)\right]-2L\cdot {\cal P}^\perp(Q)\cdot {\cal P}^\perp(L)\cdot {\cal P}^\perp(K)\cdot L\,.
\eeq
As emphasized previously, because ${\cal C}_{\kappa\lambda\tau}=0$ for $\kappa+\lambda+\tau\neq0$, the sum of shifted momenta in \eqn{eq:49} vanishes: $Q_\kappa+K_\lambda+L_\tau=0$. This allows us to use a number of manipulations similar to those used in the case of a vanishing background field.  First, we symmetrize the function ${\cal B}$ under the momentum integral in \eqn{eq:49} and use the identity
\beq
\label{appeq:symB}
 \frac{1}{6}\left\{ {\cal B}(Q,K,L)+{\rm perm.}\right\}=\frac{1}{3}\left\{\left(d-\frac{3}{2}\right)\left(\frac{1}{K^2}+\frac{1}{Q^2}+\frac{1}{L^2}\right)+\frac{K^4+L^4+Q^4}{4Q^2L^2K^2}\right\}\left[L^2 K^2-\left(L\cdot K\right)^2\right],
\eeq
where ``perm.'' denotes all possible permutations of the momenta $Q$, $K$, and $L$.
Note that, for $Q+K+L=0$, the last factor on the right-hand side is symmetric in $Q,K,L$; that is,
\beq
 L^2 K^2-\left(L\cdot K\right)^2=Q^2 K^2-\left(Q\cdot K\right)^2=L^2 Q^2-\left(L\cdot Q\right)^2.
\eeq
Next, we ``desymmetrize'' the right-hand side of \Eqn{appeq:symB} under the integral in \eqn{eq:49}, which amounts to replacing 
\beq
  {\cal B}(Q_\kappa,K_\lambda,L_\tau)\to \left\{\left(d-\frac{3}{2}\right)\frac{1}{Q_\kappa^2}+\frac{Q_\kappa^2}{4L_\tau^2K_\lambda^2}\right\}\left[L_\tau^2 K_\lambda^2-\left(L_\tau\cdot K_\lambda\right)^2\right]
\eeq
in \eqn{eq:49}. The momentum integral in \Eqn{eq:49} can thus be written
\begin{align}
&\int_{Q,K} \left[L_\tau^2K_\lambda^2-(L_\tau\cdot K_\lambda)^2\right]\left\{ \left(d-\frac{3}{2}\right)\frac{G_m(Q_\kappa)}{Q_\kappa^2}G_m(K_\lambda)G_m(L_\tau)
 +  \frac{1}{4}\left[1-m^2G_m(Q_\kappa)\right]\frac{G_m(K_\lambda)}{K_\lambda^2}\frac{G_m(L_\tau)}{L_\tau^2}\right\},
\end{align}
where we used $Q^2G_m(Q)=1-m^2G_m(Q)$. Using the trick \eqn{eq:themegatrick} as well as the symmetry properties of the tensor \eqn{eq:tensor} and of the integral \eqn{eq:integral}, we obtain \Eqn{eq:sunset}, after some simple manipulations.

%%%%%   Reduction of the integral $I^{\kappa\lambda\tau}_{\alpha\beta\gamma}$ to the scalar sunset integral $S^{\kappa\lambda\tau}_{\alpha\beta\gamma}$
\section{Reduction of the integral $I^{\kappa\lambda\tau}_{\alpha\beta\gamma}$ to the integrals $J^\kappa_\alpha$, $\tilde J^\kappa_\alpha$, and $S^{\kappa\lambda\tau}_{\alpha\beta\gamma}$}
\label{appsec:reduc}

Here, we show how the integral $I^{\kappa\lambda\tau}_{\alpha\beta\gamma}$, \Eqn{eq:integral}, can be expressed in terms of the scalar tadpole and sunset loop integrals $J_\alpha^\kappa$, $\tilde J_\alpha^\kappa$ and $S^{\kappa\lambda\tau}_{\alpha\beta\gamma}$ given in Eqs. \eqn{eq:tadpole}, \eqn{eq:tadpoleq0}, and \eqn{eq:scalarsunset}, respectively. To do so, we work out the various pieces of the integrand in \Eqn{eq:integral}. We first write 
\begin{align}
 Q^2 K^2G_\alpha(Q)G_\beta(K)&=(Q^2+\alpha^2-\alpha^2)(K^2+\beta^2-\beta^2)G_\alpha(Q)G_\beta(K)\nonumber\\
&=1-\left[\alpha^2 G_\alpha(Q)+\beta^2 G_\beta(K)\right]+\alpha^2\beta^2 G_\alpha(Q)G_\beta(K)\,,
\end{align}
and, using the identity
\beq
Q\cdot K=\frac{1}{2}\Big[\alpha^2+\beta^2-\gamma^2+L^2+\gamma^2-(Q^2+\alpha^2)-(K^2+\beta^2)\Big],
\eeq
we also have
\begin{align}
(Q\cdot K)^2G_\alpha(Q)G_\beta(K)G_\gamma(L)&=\left(\frac{\alpha^2+\beta^2-\gamma^2}{2}\right)^2G_\alpha(Q)G_\beta(K)G_\gamma(L)\nn
&+\frac{1}{2}\left(\frac{\alpha^2+\beta^2-\gamma^2}{2}+Q\cdot K\right)\Big\{G_\alpha(Q)G_\beta(K)-\left[G_\alpha(Q)+G_\beta(K)\right]G_\gamma(L)\Big\}\,.
\end{align}
Combining the above identities, we obtain
\begin{align}
\left[Q^2 K^2-\left(Q\cdot K\right)^2\right]&G_\alpha(Q)G_\beta(K)G_\gamma(L)\nn
&=\frac{1}{2}\Big\{2+Q\cdot K\left[G_\alpha(Q)+G_\beta(K)\right]\Big\}G_\gamma(L)-\frac{Q\cdot K}{2}G_\alpha(Q)G_\beta(K)\nonumber\\
& -\frac{1}{4}(\alpha^4+\beta^4+\gamma^4-2\alpha^2\beta^2-2\alpha^2\gamma^2-2\beta^2\gamma^2)G_\alpha(Q)G_\beta(K)G_\gamma(L)\nonumber\\
\label{appeq:well}
& +\frac{\gamma^2-\alpha^2-\beta^2}{4}G_\alpha(Q)G_\beta(K)+\frac{\beta^2-3\alpha^2-\gamma^2}{4}G_\alpha(Q)G_\gamma(L)+\frac{\alpha^2-3\beta^2-\gamma^2}{4} G_\beta(K)G_\gamma(L)\,.\nonumber\\
\end{align}
The term in curly brackets on the right-hand side can be rewritten as
\beq
\label{appeq:lasttrick}
 \Big[1+Q\cdot K G_\alpha(Q)\Big]+\Big[1+Q\cdot K G_\beta(K)\Big]=\left(\alpha^2-Q\cdot L\right) G_\alpha(Q)+\left(\beta^2-K\cdot L\right)G_\beta(K),
\eeq
where we used the fact that $Q+K+L=0$, which, as already emphasized, holds as well for the shifted momenta in \Eqn{eq:integral} under the average with weight ${\cal C}_{\kappa\lambda\tau}$. Inserting \eqn{appeq:lasttrick} in \eqn{appeq:well}, we obtain the following, manifestly symmetric expression for the integrand in \eqn{eq:integral}:
\begin{align}
 4\left[Q^2 K^2-\left(Q\cdot K\right)^2\right]G_\alpha(Q)G_\beta(K)G_\gamma(L)
&  = \left(2\alpha^2\beta^2+2\alpha^2\gamma^2+2\beta^2\gamma^2-\alpha^4-\beta^4-\gamma^4\right)G_\alpha(Q)G_\beta(K)G_\gamma(L)\nn
&+\left(\gamma^2-\alpha^2-\beta^2-2Q\cdot K\right)G_\alpha(Q)G_\beta(K)\nn
&+ \left(\alpha^2-\beta^2-\gamma^2-2K\cdot L\right)G_\beta(K)G_\gamma(L)\nn
&+\left(\beta^2-\gamma^2-\alpha^2-2L\cdot Q\,\right)G_\gamma(L)G_\alpha(Q).
\end{align}
It is then immediate to obtain the desired relation:
\begin{align}
I^{\kappa\lambda\tau}_{\alpha\beta\gamma} & =  \frac{1}{4}\left[\left(\gamma^2-\alpha^2-\beta^2\right)J_\alpha^\kappa J_\beta^\lambda+\left(\beta^2-\alpha^2-\gamma^2\right)J_\alpha^\kappa J_\gamma^\tau +\left(\alpha^2-\beta^2-\gamma^2\right)J_\beta^\lambda J_\gamma^\tau\right]\nonumber\\
\label{appeq:relation}
&-\frac{1}{2}\Big[\tilde J_{\alpha}^\kappa \tilde J_{\beta}^\lambda+ \tilde J_{\alpha}^\kappa \tilde J_{\gamma}^\tau+\tilde J_{\beta}^\lambda \tilde J_{\gamma}^\tau\Big] - \frac{1}{4}\left(\alpha^4+\beta^4+\gamma^4-2\alpha^2\beta^2-2\alpha^2\gamma^2-2\beta^2\gamma^2\right) S^{\kappa\lambda\tau}_{\alpha\beta\gamma}\,,
\end{align}
where we used
\beq
\int_{Q,K}\left(Q_\kappa\cdot K_\lambda\right) G_\alpha(Q_\kappa)G_\beta(K_\lambda)= \int_{Q} Q_\kappa^0 G_\alpha(Q_\kappa)\int_KK_\lambda^0G_\beta(K_\lambda)=\tilde J_{\alpha}^\kappa\tilde J_{\beta}^\lambda.
\eeq
The relation \eqn{appeq:relation} reads, explicitly, for the various cases of interest,
\bea
\label{appeq:relation1}
I^{\kappa\lambda\tau}_{mmm} & = & -\frac{1}{2}\big[\tilde J_m^\kappa \tilde J_m^\lambda+\tilde J_m^\kappa \tilde J_m^\tau+\tilde J_m^\lambda \tilde J_m^\tau\big]-\frac{m^2}{4}\big[J_m^\kappa J_m^\lambda+J_m^\kappa J_m^\tau+J_m^\lambda J_m^\tau\big]+\frac{3m^4}{4} S_{mmm}^{\kappa\lambda\tau}\\
\label{appeq:relation3}
I^{\kappa\lambda\tau}_{m00} & = & -\frac{1}{2}\big[\tilde J_m^\kappa \tilde J_0^\lambda+\tilde J_m^\kappa \tilde J_0^\tau +\tilde J_0^\lambda \tilde J_0^\tau\big]-\frac{m^2}{4}\left[J_m^\kappa J_0^\lambda+J_m^\kappa J_0^\tau-J_0^\lambda J_0^\tau\right]-\frac{m^4}{4}S^{\kappa\lambda\tau}_{m00}\\
\label{appeq:relation2}
I^{\kappa\lambda\tau}_{mm0} & = &  -\frac{1}{2}\big[\tilde J_m^\kappa \tilde J_m^\lambda + \tilde J_m^\kappa \tilde J_0^\tau+\tilde J_m^\lambda \tilde J_0^\tau\big]-\frac{m^2}{2}J_m^\kappa J_m^\lambda\\
\label{appeq:relation4}
I^{\kappa\lambda\tau}_{000} & = & -\frac{1}{2}\big[\tilde J_0^\kappa \tilde J_0^\lambda+\tilde J_0^\kappa \tilde J_0^\tau+\tilde J_0^\lambda \tilde J_0^\tau\big].
\eea
Using these relations in \Eqn{eq:touloupewithI}, we obtain \Eqn{eq:fff}.

\end{widetext}

%%%%%   Matsubara sums
\section{Matsubara sums}
\label{appsec:matsubara}

In this section, we perform explicitly the various sums over Matsubara frequencies and angular momentum integrations involved in the one- and two-loop scalar integrals derived in the previous section and we extract the UV divergent parts using dimensional regularization.  In what follows, we note $\hat r=rT$, such that shifted momenta read $Q_\kappa=(\omega_n+\kappa\hat r,\bf q\,)$.

%%%   Tadpoles
\subsection{Tadpoles}

We begin with the tadpole integral
\beq
J^\kappa_\alpha=\int_Q G_\alpha(Q_\kappa)\,.
\eeq
Note that $J_\alpha^\kappa=J_\alpha^{-\kappa}$. Standard contour integration techniques yield
\begin{align}
J^\kappa_\alpha & = \mu^{2\epsilon}\int_{\bf q}\frac{1}{2\varepsilon_{\alpha,q}}[n_{\varepsilon_{\alpha,q}-i\kappa \hat r}-n_{-\varepsilon_{\alpha,q}-i\kappa \hat r}]\nonumber\\
\label{appeq:eachterm}
& = J^{\rm vac}_{\alpha}+\int_{\bf q}\, {\rm Re}\,\frac{n_{\varepsilon_{\alpha,q}-i\kappa \hat r}}{\varepsilon_{\alpha,q}}\,,
\end{align}
where $\int_{\bf q}=\int\frac{d^{d-1}q}{(2\pi)^{d-1}}$, $\varepsilon_{\alpha,q}=\sqrt{q^2+\alpha^2}$, and $n_z=(\exp \beta z -1)^{-1}$ is the Bose-Einstein distribution function, which satisfies $n_{-x}=-1-n_x$. Here, we extracted explicitly a zero temperature, background-field-independent contribution $J^{\rm vac}_{\alpha}$. For later use, we introduce the following notation
\beq
\label{appeq:J0n1n}
J^\kappa_\alpha=J_\alpha(0n)+J^\kappa_\alpha(1n),
\eeq
which emphasizes the number of thermal factors (i.e., Bose-Einstein distribution functions) appearing in each term on the right-hand side of \eqn{appeq:eachterm}. In particular, $J_\alpha(0n)=J^{\rm vac}_{\alpha}$. In dimensional regularization $J_0(0n)=0$ and
\begin{align}
\label{eq:taddiv}
J_m(0n)&= \mu^{2\epsilon}\int_{\bf q}\frac{1}{2\varepsilon_q}= \mu^{2\epsilon}\int\frac{d^dQ}{(2\pi)^d}\frac{1}{Q^2+m^2}\nn
&=-\frac{m^2}{16\pi^2}\left[\frac{1}{\epsilon}+\ln\frac{\bar\mu^2}{m^2}+1+{\cal O}(\epsilon)\right],
\end{align}
where $\bar\mu^2=4\pi e^{-\gamma}\mu^2$, with $\gamma$ the Euler constant. 

The other tadpole integral which appears in \Eqn{eq:fff} is
\beq
\tilde J^\kappa_{\alpha}=\int_Q Q^0_\kappa G_\alpha(Q_\kappa).
\eeq
The Matsubara sum in this expression is not absolutely convergent but it can be defined as the limit of the symmetric summation $\sum_{n=-N}^N$ for $N\to\infty$. Alternatively, we can make it an absolutely convergent sum by writing 
\begin{align}
\sum_{n=-N}^N Q_\kappa^0 G_\alpha(Q_\kappa)= \sum_{n=-N}^N \Big[Q_\kappa^0 G_\alpha(Q_\kappa)-Q^0 G_\alpha(Q)\Big],
\end{align}
where the added term vanishes by symmetry. We can then use standard contour techniques to obtain
\begin{align}
\tilde J^\kappa_{\alpha} &= \int_{\bf q}\frac{1}{2i}\Big[n_{\varepsilon_{\alpha,q}-i\kappa \hat r}+n_{-\varepsilon_{\alpha,q}-i\kappa \hat r}-n_{\varepsilon_{\alpha,q}}-n_{-\varepsilon_{\alpha,q}}\Big]\nonumber\\
\label{appeq:poerr}
 & = \int_{\bf q} \,{\rm Im}\,n_{\varepsilon_{\alpha,q}-i\kappa \hat r}\,,
\end{align}
which satisfies $\tilde J^{-\kappa}_\alpha=-\tilde J^\kappa_\alpha$ and, in particular, $\tilde J_\alpha^0=0$. The zero temperature contribution vanishes identically, $\tilde J^\kappa_{\alpha} (0n)=0$, and we also note that $\tilde J^\kappa_{\alpha}=0$ when $r$ is a multiple of $\pi$. The integral \eqn{appeq:poerr} is finite and never multiplies a divergent contribution, so we can safely set $d\to4$ there.

%%%   The scalar sunset
\subsection{The scalar sunset}

We now treat the scalar two-loop integral
\beq
\label{appeq:scalarsunset}
S^{\kappa\lambda\tau}_{\alpha\beta\gamma}  \equiv  \int_{Q,K} G_\alpha(Q_\kappa)G_\beta(K_\lambda)G_\gamma(L_\tau)\,,
\eeq
where $Q+K+L=0$, for the relevant case of conserved shifts, $\kappa+\lambda+\tau=0$, which implies $Q_\kappa+K_\lambda+L_\tau=0$. We extend the approach of \cite{Blaizot:2004bg}. It proves useful to introduce the spectral representation of the (free) propagators
\beq
\label{appeq:GGdef}
G_\alpha(Q_\kappa)=\tilde G_\alpha(i\omega_n^\kappa;q)\equiv\int_{q_0}\frac{\rho_\alpha(q_0,q)}{q_0-i\omega_n^\kappa},
\eeq
where $\int_{q_0}=\int dq_0/(2\pi)$, $\omega_n^\kappa=\omega_n+\kappa \hat r$ is the shifted Matsubara frequency, and
\beq
 \rho_\alpha(q_0,q)=2\pi\,{\rm sign}(q_0)\delta\!\left(q^2_0-\varepsilon_{\alpha,q}^2\right),
\eeq
with $\varepsilon_{\alpha,q}=\sqrt{q^2+\alpha^2}$. The double Matsubara sum in \eqn{appeq:scalarsunset} yields
\begin{align}
& T^2\sum_{n,m}\frac{1}{(q_0-i\omega_n^\kappa)(k_0-i\omega_m^\lambda)(l_0+i\omega_n^\kappa+i\omega_m^\lambda)}\,.\nonumber\\
& \hspace{.5cm} =\,\frac{(n_{k_0-i\lambda \hat r}-n_{-l_0+i\tau \hat r})(n_{q_0-i\kappa \hat r}-n_{-l_0-k_0-i\kappa \hat r})}{l_0+k_0+q_0}\nonumber\\
\label{appeq:doublesum}
&  \hspace{.5cm} =\,\frac{n_{k_0-i\lambda \hat r}n_{l_0-i\tau \hat r}-n_{-q_0+i\kappa \hat r}(n_{k_0-i\lambda \hat r}-n_{-l_0+i\tau \hat r})}{l_0+k_0+q_0},
\end{align}
where we have used the identity $(1+n_x+n_y)n_{x+y}=n_x n_y$ and $\kappa+\lambda+\tau=0$. The second line of this equation makes it clear that \eqn{appeq:doublesum} is well defined for all $k_0$, $q_0$, and $l_0$, including the limiting case $l_0+k_0+q_0\to0$, for which both the numerator and the denominator vanish linearly. In the following, we shall decompose the fraction in the third line of \eqn{appeq:doublesum} in different pieces whose numerators do not vanish at $l_0+k_0+q_0=0$, thus making the corresponding contribution to \eqn{appeq:doublesum} formally divergent in this limit. To avoid this problem, we regulate the denominator in \eqn{appeq:doublesum} as
\beq
\frac{1}{ l_0+k_0+q_0}\to {\rm Re}\left(\frac{1}{l_0+k_0+q_0+i0^+}\right).
\eeq

Now, we use the identities
\begin{align}
 n_{k_0-i\kappa\hat r}&=-\theta(-k_0)+{\rm sign}(k_0)n_{|k_0|-i\,{\rm sign}(k_0) \kappa\hat r}\\
 n_{-k_0+i \kappa\hat r}&=-\theta(k_0)-{\rm sign}(k_0)n_{|k_0|-i\,{\rm sign}(k_0)\kappa\hat r}
\end{align}  
in the third line of \eqn{appeq:doublesum} to rewrite \Eqn{appeq:scalarsunset} as
\begin{align}
S_{\alpha\beta\gamma}^{\kappa\lambda\tau}  &= S_{\alpha\beta\gamma}^{\rm vac} + \mu^{2\epsilon}\int_{q_0,\bf q}\!\!\sigma_\alpha^\kappa(q_0,q)\,{\rm Re}\,\tilde I_{\beta\gamma}(q_0+i0^+;q)\nonumber\\
& + \int_{q_0,k_0,\bf q,\bf k}\sigma_\alpha^\kappa(q_0,q)\sigma_\beta^\lambda(k_0,k)\,{\rm Re}\,\tilde G_\gamma(\ell_0+i0^+;\ell)\nonumber\\
\label{appeq:inter}
&+ {\rm perm.},
\end{align}
where $S_{\alpha\beta\gamma}^{\rm vac}$ is an unimportant vacuum contribution, independent of the temperature and of the background field, $\ell_0=q_0+k_0$, $\ell=|\bf{q}+\bf{k}|$, 
\beq
 \sigma_\alpha^\kappa(q_0,q)=\rho_\alpha(q_0,q)\,{\rm sign}(q_0)\,n_{|q_0|-i\,{\rm sign}(q_0)\kappa \hat r},
\eeq
and ``perm.'' denotes the circular permutations of the pairs of indices $(\alpha,\kappa)$, $(\beta,\lambda)$, and $(\gamma,\tau)$ in the two integrals that appear explicitly in (\ref{appeq:inter}). The function $\tilde G_\alpha(z;q)$ has been defined in \eqn{appeq:GGdef} and the function $\tilde I_{\alpha\beta}(z;q)$ is related to the vacuum one-loop integral [here $Q=(\omega,\bf q)$]
\begin{align}
I^{\rm vac}_{\alpha\beta}(Q)&=\tilde I_{\alpha\beta}(i\omega;q)\equiv\mu^{2\epsilon}\!\!\int\frac{d^dK}{(2\pi)^d}G_\alpha(K)G_\beta(Q+K)\nn
\label{appeq:oneloopvac}
&=\mu^{2\epsilon}\!\!\int_{k_0,l_0,\bf k}\rho_\alpha(k_0,k)\rho_\beta(l_0,\ell)\,\frac{\theta(l_0)-\theta(-k_0)}{i\omega+l_0+k_0}.
\end{align}
In obtaining \Eqn{appeq:inter}, we have used $\tilde G_\alpha(x;q)=\tilde G_\alpha(-x;q)$ and $\tilde I_{\alpha\beta}(x;q)=\tilde I_{\alpha\beta}(-x;q)$, and we have set $d\to4$ in the second, UV finite line. In contrast, one has to keep $d$ arbitrary in the second term on the right-hand side of \Eqn{appeq:inter} since it contains UV divergent contributions, arising from the zero temperature loop \eqn{appeq:oneloopvac}.

It is now an easy matter to perform explicitly the frequency and, for the double integral on the second line, the angular integrations. As before, we decompose the result according to the number of thermal factors $n$ in each contribution as
\beq
 S_{\alpha\beta\gamma}^{\kappa\lambda\tau}=S_{\alpha\beta\gamma}^{\kappa\lambda\tau}(0n)+S_{\alpha\beta\gamma}^{\kappa\lambda\tau}(1n)+S_{\alpha\beta\gamma}^{\kappa\lambda\tau}(2n).
\eeq
We obtain
\begin{widetext}
\begin{align}
\label{appeq:S1n}
S_{\alpha\beta\gamma}^{\kappa\lambda\tau}(1n) & = \mu^{2\epsilon}\int_{\bf q}\,{\rm Re}\,\frac{n_{\varepsilon_{\alpha,q}-i\kappa \hat r}}{\varepsilon_{\alpha,q}}\,{\rm Re}\,\tilde I_{\beta\gamma}(\varepsilon_{\alpha,q}+i0^+;q)+{\rm perm.}\\
S_{\alpha\beta\gamma}^{\kappa\lambda\tau}(2n) & = \frac{1}{32\pi^4}\int_0^\infty \!\!dq\,q\int_0^\infty \!\!dk\,k\, {\rm Re}\,\frac{n_{\varepsilon_{\alpha,q}-i\kappa \hat r}\,n_{\varepsilon_{\beta,k}-i\lambda \hat r}}{\varepsilon_{\alpha,q}\,\varepsilon_{\beta,k}}\,{\rm Re}\,\ln\frac{(\varepsilon_{\alpha,q}+\varepsilon_{\beta,k}+i0^+)^2-(\varepsilon_{\gamma,k+q})^2}{(\varepsilon_{\alpha,q}+\varepsilon_{\beta,k}+i0^+)^2-(\varepsilon_{\gamma,k-q})^2}\nonumber\\
& + \frac{1}{32\pi^4}\int_0^\infty \!\!dq\,q\int_0^\infty \!\!dk\,k \,{\rm Re}\,\frac{n_{\varepsilon_{\alpha,q}-i\kappa \hat r}\,n_{\varepsilon_{\beta,k}+i\lambda \hat r}}{\varepsilon_{\alpha,q}\,\varepsilon_{\beta,k}}\,{\rm Re}\,\ln\frac{(\varepsilon_{\alpha,q}-\varepsilon_{\beta,k}+i0^+)^2-(\varepsilon_{\gamma,k+q})^2}{(\varepsilon_{\alpha,q}-\varepsilon_{\beta,k}+i0^+)^2-(\varepsilon_{\gamma,k-q})^2}\,.\nonumber\\
\label{appeq:S2n}
&+{\rm perm.},
\end{align}
\end{widetext}
and $S_{\alpha\beta\gamma}^{\kappa\lambda\tau}(0n)=S_{\alpha\beta\gamma}^{\rm vac}$. As mentioned before, the contribution $S_{\alpha\beta\gamma}^{\kappa\lambda\tau}(2n)$ is UV finite but the contribution $S_{\alpha\beta\gamma}^{\kappa\lambda\tau}(1n)$ contains UV divergent terms which explicitly depend on the temperature and on the background field. We shall check in Appendix~\ref{appsec:final} that such contributions cancel after renormalization. For this purpose, it is useful to note that \eqn{appeq:S1n} rewrites
\beq
\label{eq:nom}
S_{\alpha\beta\gamma}^{\kappa\lambda\tau}(1n)=J_\alpha^\kappa(1n)\,\,{\rm Re}\,\tilde I_{\beta\gamma}(\varepsilon_{\alpha,q}+i0^+;q)+{\rm perm.},
\eeq
where $J_\alpha^\kappa(1n)$ is defined\footnote{Strictly speaking, we should use the expression of $J(1n)$ for arbitrary $d$ in \eqn{eq:nom} since it multiplies a divergent integral. However, we readily see from \Eqn{eq:fff} that the counterterm contribution which cancels this divergence is of the form $J(1n)/\epsilon$. We thus do not need the explicit expression of $J(1n)$ to discuss renormalization and moreover, after renormalization has been performed we can take $d=4$ in order to evaluate $J(1n)$.} in \eqn{appeq:eachterm} and \eqn{appeq:J0n1n} and where we used the fact, owing to the O($d$) invariance of the Euclidean integral \eqn{appeq:oneloopvac}, ${\rm Re}\,\tilde I_{\beta\gamma}(\varepsilon_{\alpha,q}+i0^+;q)$ depends only on $\varepsilon_{\alpha,q}^2-q^2=\alpha^2$. The expression for $I^{\rm vac}_{\alpha\beta}(Q)=\tilde I_{\alpha\beta}(i\omega;q)$ can be found for instance in \cite{Reinosa:2013twa} and reads, up to corrections of ${\cal O}(\epsilon)$,\footnote{A typo in the formula (B2) of Ref. \cite{Reinosa:2013twa} is corrected here: the first three terms on the right-hand side receive a factor $1/2$ due to the symmetrization $(\alpha\leftrightarrow\beta)$.}
\begin{align}
 I^{\rm vac}_{\alpha\beta}(Q) &=\frac{1}{16\pi^2}\Bigg\{\frac{1}{2\epsilon}+1+\frac{1}{2}\ln\frac{\bar\mu^2}{Q^2}-\frac{1}{2}\ln\!\left({\cal C}^2_{\alpha\beta}(Q)-{1\over4}\right)\nn
 &+{\cal C}_{\alpha\beta}(Q)\ln\frac{{\cal C}_{\alpha\beta}(Q)-{1\over2}}{{\cal C}_{\alpha\beta}(Q)+{1\over2}}\Bigg\}+(\alpha\leftrightarrow\beta),
\end{align}
where ${\cal C}_{\alpha\beta}(Q)$ (which should not be mistaken with ${\cal C}_{\kappa\lambda\tau}$) is given by
\beq
 {\cal C}_{\alpha\beta}(Q)=\frac{B_{\alpha\beta}(Q)+\alpha^2-\beta^2}{2Q^2},
\eeq
with
\beq
 B_{\alpha\beta}(Q)=\sqrt{Q^4+2Q^2\left(\alpha^2+\beta^2\right)+\left(\alpha^2-\beta^2\right)^2}.
\eeq
Using this formula and the definition \eqn{appeq:oneloopvac}, we get
\begin{align}
\label{appeq:Idiv00}
 {\rm Re}\,\tilde I_{00}(\varepsilon_{q}\!+\!i0^+;q)&=\frac{1}{16\pi^2}\!\!\left[\frac{1}{\epsilon}+\ln\frac{\bar\mu^2}{m^2}+2\right]\\
\label{appeq:Idivm0}
 {\rm Re}\,\tilde I_{m0}({q}\!+\!i0^+;q)&=\frac{1}{16\pi^2}\!\!\left[\frac{1}{\epsilon}+\ln\frac{\bar\mu^2}{m^2}+1\right]\\
\label{appeq:Idivmm}
 {\rm Re}\,\tilde I_{mm}(\varepsilon_{q}\!+\!i0^+;q)&=\frac{1}{16\pi^2}\!\!\left[\frac{1}{\epsilon}+\ln\frac{\bar\mu^2}{m^2}+2-\frac{\pi}{\sqrt{3}}\right].
\end{align}

%%%
\subsection{Massless integrals at finite temperature}\label{appsec:massless}

When discussing the (formal) large temperature $T\gg m$ behavior of our results, we encounter the following integrals
\begin{align}
P_{2n+1}(r) & \equiv \int_0^\infty dx\,x^{2n}\,{\rm Im}\,\tilde n_{x-ir}\label{eq:def1}\\
P_{2n+2}(r) & \equiv \int_0^\infty dx\,x^{2n+1}\,{\rm Re}\,\tilde n_{x-ir}\,,\label{eq:def2}
\end{align}
with $n\in\mathds{N}$ and $\tilde n_z=1/(e^z-1)$. For $n\geq 0$, we have
\begin{align}
P_{2n+2}'(r)& = -(2n+1)P_{2n+1}(r),\\
P_{2n+3}'(r)&=(2n+2) P_{2n+2}(r).
\end{align}
It follows that
\beq\label{appeq:dn}
P_{n}^{(n-1)}(r)=(-1)^{\left \lfloor{n/2}\right \rfloor}(n-1)!\,P_1(r)\,,
\eeq
where $\left \lfloor x\right \rfloor={\rm max}\{n\in\mathds{N}\,|\,n\leq x\}$ is the integer part of $x$. In order to obtain $P_n(r)$ for any $n>1$ from the knowledge of $P_1(r)$, Eq.~(\ref{appeq:dn}) has to be supplemented by the conditions  $P_{2n+1}(\pi)=0$ and $P_{2n+2}(0)=(2n+1)!\zeta(2n+2)$ which are easily checked from the definitions (\ref{eq:def1}) and (\ref{eq:def2}).\footnote{We also have $P_{2n+1}(0)=0$ for $n\geq 1$ but $P_1(0)=\pi/2$. This is because, even though the integrand of $P_1$ goes to $0$ as $r\to 0$, the integral behaves like $$\frac{1}{2}\sin(r)\int_0^\infty \frac{dx}{\cosh(x)-\cos(r)}\sim \sin(r)\int_0^\infty \frac{dx}{x^2+r^2}\to\frac{\pi}{2}\,.$$}

To compute $P_1(r)$, we note that, upon expanding the Bose-Einstein factor $\tilde n_z$ as a geometric series, $\tilde n_z=\sum_{k=0}^\infty e^{-(k+1)z}$, it can be rewritten as
\beq
P_1(r)=\sum_{k=-\infty}^\infty \frac{e^{ikr}}{2ik}(1-\delta_{k0})\,,\\
\eeq
a Fourier series whose sum is nothing but $P_1(r)=(\pi-r)/2$ in the interval $]0,2\pi[$. This implies in particular that the $P_n(r)$'s are polynomials in the interval $]0,2\pi[$ and later we shall need the first of them:
\begin{align}
P_1(r) & = \frac{\pi-r}{2}\\
P_2(r) & = \frac{(\pi-r)^2}{4}-\frac{\pi^2}{12}\\
P_3(r) & = -\frac{(\pi-r)^3}{6}+\frac{\pi^2 (\pi-r)}{6}\\
P_4(r) & = -\frac{(\pi-r)^4}{8}+\frac{\pi^2 (\pi-r)^2}{4}-\frac{7\pi^4}{120}\,.\label{appeq:P4}
\end{align}

%%%%%   Final expression of $\gamma^{(2)}$
\section{Final expression of $V^{(2)}(T,r)$}
\label{appsec:final}

We put together the material derived in the previous sections to obtain a final, explicitly finite expression for the two-loop contribution to the background field potential in terms of one- and two-dimensional (radial) momentum integrals. 

%%%   Renormalization
\subsection{Renormalization}
\label{appsec:renorm}

We perform the renormalization at $T=0$ in which case $\bar A_{\rm min}=0$ and the Landau-DeWitt gauge coincides with the Landau gauge.  The mass counterterm $\delta Z_{m^2}$ has been computed in \cite{Kondo:2001tm,Gracey:2002yt} and reads
\beq
 \delta Z_{m^2}=\frac{g^2N}{192\pi^2}\left[-\frac{35}{\epsilon}+z_f\right],
\eeq
where the finite part $z_f$ depends on the renormalization scheme. With the renormalization conditions (\ref{eq:ren_cond}), one obtains \cite{Tissier_10,Tissier_11} 
\begin{widetext}
\bea
z_f & = & -\frac{1}{s^2}+\frac{111}{
2s}-\frac{287}{6}-35\ln \left(\bar s \right)-\frac{1}{2}\left(s^2-2\right) \ln (s)\nonumber\\
& & +\left(s^2-10 s+1\right) \left(\frac{1}{s}+1\right)^3 \ln
   (s+1)+\, \frac{1}{2}(s^2-20s+12)\left(\frac{4}{s}+1\right)^{3/2}  \ln
\left(\frac{\sqrt{4/s+1}-1}{\sqrt{4/s+1}+1}\right),
\eea
\end{widetext}
with $s\equiv \mu^2/m^2$ and $\bar s\equiv \bar\mu^2/m^2$. Using Eqs. \eqn{eq:fff}, \eqn{eq:nom}, \eqn{appeq:Idiv00}--\eqn{appeq:Idivmm}, and \eqn{eq:contraint2}, one explicitly checks that the temperature- and background-field-dependent divergences cancel out. That no divergences are generated by the background field can be understood from the background gauge invariance [see \Eqn{eq_ginv}] as we shall discuss in a future work. Setting $d=4$ in the remaining finite expression, we obtain, for the thermal contribution,
\begin{align}
V^{(2)} (T,r)& = m^2 C_N \sum_\kappa J_m^\kappa(1n)\nonumber\\
& +  \frac{3g^2}{8}\sum_{\kappa,\lambda,\tau}{\cal C}_{\kappa\lambda\tau}\left[\frac{5}{2}U^\kappa V^\lambda- \frac{7}{m^2}\tilde U^\kappa\tilde V^\lambda\right]\nonumber\\
& +  \frac{g^2m^2}{16}\sum_{\kappa,\lambda,\tau}{\cal C}_{\kappa\lambda\tau}\left[33S_{mmm}^{\kappa\lambda\tau}(2n)+S_{m00}^{\kappa\lambda\tau}(2n)\right],
\end{align}
where we used the symmetry of the tensor ${\cal C}_{\kappa\lambda\tau}$ and we defined\footnote{We note that the explicit $\ln \bar s$ cancels against the one in $z_f$.} 
\beq
C_N=\frac{g^2N}{128\pi^2}\left(z_f+35\ln\bar s+\frac{313}{3}-\frac{99\pi}{2\sqrt{3}}\right)
\eeq
and
\begin{align}
 U^\kappa&=J_m^\kappa(1n)+\frac{1}{3}J^\kappa_0(1n)\\
 V^\kappa&=J_m^\kappa(1n)-\frac{1}{5}J^\kappa_0(1n)\\
 \tilde U^\kappa&=\tilde J_{m}^\kappa-\tilde J_{0}^\kappa\\
 \tilde V^\kappa&=\tilde J_{m}^\kappa-\frac{5}{21}\tilde J_{0}^\kappa.
\end{align}

Finally, we emphasize that the results derived so far do not rely on the explicit form of the tensor ${\cal C}_{\kappa\lambda\tau}$ and only use the conservation of color charge (and hence of shifted momenta) at the interaction vertices. All formulas thus hold for a general group SU($N$). Note that, in general, $r$ must be understood as a vector in the $(N-1)$--dimensional Cartan subalgebra.

%%%%%
\subsection{Finite contributions}

Here, we specify to the case $N=2$ and use the values of the tensor ${\cal C}_{\kappa\lambda\tau}$ to obtain a more explicit formula for $V^{(2)}(T,r)$. We have, for any quantity ${\cal F}^{\kappa\lambda\tau}$,
\begin{align}
\sum_{\kappa,\lambda,\tau} {\cal C}_{\kappa\lambda\tau}{\cal F}^{\kappa\lambda\tau} &=  {\cal F}^{0+-}+{\cal F}^{+-0}+{\cal F}^{-0+}\nn
&+{\cal F}^{0-+}+{\cal F}^{-+0}+{\cal F}^{+0-}.
\end{align}
Using \Eqn{appeq:S2n}, we obtain
\begin{widetext}
\begin{align}
 \sum_{\kappa,\lambda,\tau} {\cal C}_{\kappa\lambda\tau}S^{\kappa\lambda\tau}_{\alpha\beta\gamma}(2n)&=\frac{1}{16\pi^4}\int_0^\infty \!\!dq\,\frac{q}{\varepsilon_{\alpha,q}}\int_0^\infty\!\!dk\,\frac{k}{\varepsilon_{\beta,k}}\nonumber\\
\times \Bigg\{&{\rm Re}\left(n_{\varepsilon_{\alpha,q}-i \hat r}n_{\varepsilon_{\beta,k}+i \hat r}+n_{\varepsilon_{\alpha,q}+i \hat r}n_{\varepsilon_{\beta,k}}+n_{\varepsilon_{\alpha,q}}n_{\varepsilon_{\beta,k}-i \hat r}\right)\,{\rm Re}\,\ln\frac{(\varepsilon_{\alpha,q}+\varepsilon_{\beta,k}+i0^+)^2-(\varepsilon_{\gamma,k+q})^2}{(\varepsilon_{\alpha,q}+\varepsilon_{\beta,k}+i0^+)^2-(\varepsilon_{\gamma,k-q})^2}\nonumber\\
+\,\,&{\rm Re}\left(n_{\varepsilon_{\alpha,q}-i \hat r}n_{\varepsilon_{\beta,k}-i \hat r}+n_{\varepsilon_{\alpha,q}+i \hat r}n_{\varepsilon_{\beta,k}}+n_{\varepsilon_{\alpha,q}}n_{\varepsilon_{\beta,k}+i \hat r}\right)\,{\rm Re}\,\ln\frac{(\varepsilon_{\alpha,q}-\varepsilon_{\beta,k}+i0^+)^2-(\varepsilon_{\gamma,k+q})^2}{(\varepsilon_{\alpha,q}-\varepsilon_{\beta,k}+i0^+)^2-(\varepsilon_{\gamma,k-q})^2}\Bigg\}\nn
\label{appeq:olala}
+\,\,&{\rm perm.},
\end{align}
where ``perm.'' denotes the cyclic permutations of $(\alpha,\beta,\gamma)$. It is convenient to reorganize the Bose-Einstein distribution functions in such a way that only one contains a complex argument. Using the identities $n_x n_y=n_{x+y}(1+n_x+n_y)=n_{x-y}n_y+n_{y-x}n_x$, we have
\begin{align}\label{eq:bose}
{\rm Re}\left(n_{\varepsilon_{\alpha,q}-i \hat r}n_{\varepsilon_{\beta,k}+i \hat r}+n_{\varepsilon_{\alpha,q}+i \hat r}n_{\varepsilon_{\beta,k}}+n_{\varepsilon_{\alpha,q}}n_{\varepsilon_{\beta,k}-i \hat r}\right)&=n_{\varepsilon_{\alpha,q}+\varepsilon_{\beta,k}}\nn
&+\left(n_{\varepsilon_{\alpha,q}+\varepsilon_{\beta,k}}+n_{\varepsilon_{\beta,k}}\right){\rm Re}\,n_{\varepsilon_{\alpha,q}-i \hat r}\nn
&+\left(n_{\varepsilon_{\alpha,q}+\varepsilon_{\beta,k}}+n_{\varepsilon_{\alpha,q}}\right){\rm Re}\,n_{\varepsilon_{\beta,k}-i \hat r}\\\nn
{\rm Re}\left(n_{\varepsilon_{\alpha,q}-i \hat r}n_{\varepsilon_{\beta,k}-i \hat r}+n_{\varepsilon_{\alpha,q}+i \hat r}n_{\varepsilon_{\beta,k}}+n_{\varepsilon_{\alpha,q}}n_{\varepsilon_{\beta,k}+i \hat r}\right)&=\left(n_{\varepsilon_{\beta,k}-\varepsilon_{\alpha,q}}+n_{\varepsilon_{\beta,k}}\right){\rm Re}\,n_{\varepsilon_{\alpha,q}-i \hat r}\nn
&+\left(n_{\varepsilon_{\alpha,q}-\varepsilon_{\beta,k}}+n_{\varepsilon_{\alpha,q}}\right){\rm Re}\,n_{\varepsilon_{\beta,k}-i \hat r}\label{eq:bose2}
\end{align}
Putting everything together, we finally obtain 
\bea
V^{(2)}(T,r) & = & m^2 C_2 \left[J_m^0(1n)+2J_m^+(1n)\right]+ \frac{15g^2}{8}\left(U^0 V^++U^+V^0+U^+V^+\right)+\frac{21g^2}{4m^2}\tilde U^+ \tilde V^+\nonumber\\
& + & \frac{99g^2m^2}{256\pi^4}\int_0^\infty dq\,\frac{q}{\varepsilon_q}\int_0^\infty dk\,\frac{k}{\varepsilon_k} \Big[n_{\varepsilon_q+\varepsilon_k}+2(n_{\varepsilon_q+\varepsilon_k}+n_{\varepsilon_k}){\rm Re}\,n_{\varepsilon_q-i \hat r}\Big]\nonumber\\
& & \hspace{6.0cm}\times\,{\rm Re}\,\ln\frac{(\varepsilon_q+\varepsilon_k+i0^+)^2-(\varepsilon_{k+q})^2}{(\varepsilon_q+\varepsilon_k+i0^+)^2-(\varepsilon_{k-q})^2}\nonumber\\
& + & \frac{99g^2m^2}{256\pi^4}\int_0^\infty dq\,\frac{q}{\varepsilon_q}\int_0^\infty dk\,\frac{k}{\varepsilon_k} \Big[2(n_{\varepsilon_k-\varepsilon_q}+n_{\varepsilon_k}){\rm Re}\,n_{\varepsilon_q-i \hat r}\Big]\nonumber\\
& & \hspace{6.0cm}\times\,{\rm Re}\,\ln\frac{(\varepsilon_q-\varepsilon_k+i0^+)^2-(\varepsilon_{k+q})^2}{(\varepsilon_q-\varepsilon_k+i0^+)^2-(\varepsilon_{k-q})^2}\nonumber\\
& + & \frac{g^2m^2}{256\pi^4}\int_0^\infty dq\,\int_0^\infty dk\,\Big[n_{q+k}+2(n_{q+k}+n_{k}){\rm Re}\,n_{q-i \hat r}\Big]\nonumber\\
& & \hspace{6.0cm}\times\,{\rm Re}\,\ln\frac{(q+k+i0^+)^2-(\varepsilon_{k+q})^2}{(q+k+i0^+)^2-(\varepsilon_{k-q})^2}\nonumber\\
& + & \frac{g^2m^2}{256\pi^4}\int_0^\infty dq\,\int_0^\infty dk\,\Big[2(n_{k-q}+n_{k}){\rm Re}\,n_{q-i \hat r}\Big]\nonumber\\
& & \hspace{6.0cm}\times\,{\rm Re}\,\ln\frac{(q-k+i0^+)^2-(\varepsilon_{k+q})^2}{(q-k+i0^+)^2-(\varepsilon_{k-q})^2}\nonumber\\
& + & \frac{g^2m^2}{128\pi^4}\int_0^\infty dq\,\frac{q}{\varepsilon_q}\int_0^\infty dk\,\Big[n_{\varepsilon_q+k}+(n_{\varepsilon_q+k}+n_{k}){\rm Re}\,n_{\varepsilon_q-i \hat r}+(n_{\varepsilon_q+k}+n_{\varepsilon_q}){\rm Re}\,n_{k-i \hat r}\Big]\nonumber\\
& & \hspace{7.0cm}\times\,{\rm Re}\,\ln\frac{(\varepsilon_q+k+i0^+)^2-(k+q)^2}{(\varepsilon_q+k+i0^+)^2-(k-q)^2}\nonumber\\
& + & \frac{g^2m^2}{128\pi^4}\int_0^\infty dq\,\frac{q}{\varepsilon_q}\int_0^\infty dk\,\Big[(n_{k-\varepsilon_q}+n_{k}){\rm Re}\,n_{\varepsilon_q-i \hat r}+(n_{\varepsilon_q-k}+n_{\varepsilon_q}){\rm Re}\,n_{k-i \hat r}\Big]\nonumber\\
\label{appeq:final}
& & \hspace{7.0cm}\times\,{\rm Re}\,\ln\frac{(\varepsilon_q-k+i0^+)^2-(k+q)^2}{(\varepsilon_q-k+i0^+)^2-(k-q)^2}\,,
\eea
\end{widetext}
where we used the fact that $J_\alpha^+(1n)=J_\alpha^-(1n)$,  $\tilde J_\alpha^+(1n)=-\tilde J_\alpha^-(1n)$, and $\tilde J_\alpha^0(1n)=0$, which implies that $U^+=U^-$, $V^+=V^-$, $\tilde U^+=-\tilde U^-$, $\tilde V^+=-\tilde V^-$, and $\tilde U^0=\tilde V^0=0$. We recall that the relevant tadpole integrals are given in Eqs. \eqn{appeq:eachterm} and \eqn{appeq:poerr}. Finally, one has
\begin{align}
{\rm Re}\, n_{\varepsilon-i \hat r}&=\frac{e^{\beta\varepsilon}\cos r-1}{e^{2\beta\varepsilon}-2e^{\beta\varepsilon}\cos r+1}\\
\label{appeq:imn}
 {\rm Im}\, n_{\varepsilon-i \hat r}&=\frac{e^{\beta\varepsilon}\sin r}{e^{2\beta\varepsilon}-2e^{\beta\varepsilon}\cos r+1}.
\end{align}

%%%%%
\subsection{High and low temperature behavior of $V(T,r)$}\label{appsec:rinf}

Let us first show that, at low temperatures, $V(T,r)/T^4$ does not receive any two-loop contribution. Indeed, up to exponentially suppressed terms, we have
\begin{widetext}
\begin{align}
\frac{V^{(2)}(T,r)}{T^4} & = -\frac{g^2}{32\pi^4}\int_0^\infty dx\,x\, (2\tilde n_x+{\rm Re}\,\tilde n_{x-ir})\int_0^\infty dy\,y\,{\rm Re}\,\tilde n_{y-ir}+\frac{5g^2T^2}{16m^2\pi^4}\left(\int_0^\infty dx\,x^2\,{\rm Im}\,\tilde n_{x-ir}\right)^2\nonumber\\
& + \frac{g^2m^2}{256\pi^4T^2}\int_0^\infty dx\,\int_0^\infty dy\,\Big[\tilde n_{x+y}+2(\tilde n_{x+y}+\tilde n_{y}){\rm Re}\,\tilde n_{x-ir}\Big]{\rm Re}\,\ln\frac{(x+y+i0^+)^2-(\tilde\varepsilon_{x+y})^2}{(x+y+i0^+)^2-(\tilde\varepsilon_{x-y})^2}\nonumber\\
& + \frac{g^2m^2}{256\pi^4 T^2}\int_0^\infty dx\,\int_0^\infty dy\,\Big[2(\tilde n_{x-y}+\tilde n_{y}){\rm Re}\,\tilde n_{x-i r}\Big]{\rm Re}\,\ln\frac{(x-y+i0^+)^2-(\tilde\varepsilon_{x+y})^2}{(x-y+i0^+)^2-(\tilde\varepsilon_{x-y})^2}\nonumber\\
& + {\cal O}(e^{-m/T})\,,
\end{align}
\end{widetext}
where we have rescaled the integration variables as $q=Tx$ and $k=Ty$ and we have introduced $\tilde n_{x-ir}=n_{q-i\hat r}$ and $(\tilde\varepsilon_{x\pm y})^2=(x\pm y)^2+m^2/T^2$. In the limit $T\to 0$, each logarithm contributes $4xyT^2/m^2$. We can thus combine the two integrals and, after some simple manipulations that undo (\ref{eq:bose}) and (\ref{eq:bose2}), we arrive at
\beq
\lim_{T\to 0}\frac{V^{(2)}(T,r)}{T^4}=0\,.
\eeq
This shows, in particular, that the two-loop corrections to the background field potential do not yield any $T^4$ contribution to the pressure at low temperatures.

To discuss the high temperature behavior, we note that the only contributions that survive in this regime, of order $T^4$, all come from the first line of (\ref{appeq:final}). One obtains
\begin{align}
\label{appeq:vinf}
v_\infty(r) & \equiv \lim_{T\to\infty}\frac{V(T,r)}{T^4}\nonumber\\
& =  -\frac{1}{3\pi^2}\int_0^\infty \!\!\!dx\,x^3 \big(\tilde n_x+2{\rm Re}\,\tilde n_{x-ir}\big)\nonumber\\
& \!\!\!\!\!\!\!\!+ \frac{g^2}{2\pi^4}\int_0^\infty \!\!\!dx\,x\,\big({\rm Re}\,\tilde n_{x-ir}+2\tilde n_x\big)\int_0^\infty \!\!\!dy\,y\,{\rm Re}\,\tilde n_{y-ir}\nonumber\\
& \!\!\!\!\!\!\!\!+ \frac{g^2}{\pi^4}\int_0^\infty \!\!\!dx\,x^2\,{\rm Im}\left.\frac{\partial \tilde n_{\sqrt{x^2+z}-ir}}{\partial z}\right|_{z=0}\int_0^\infty \!\!\!dy\,y^2\,{\rm Im}\,\tilde n_{y-ir}.
\end{align}
The first line is nothing but the Weiss potential ${\cal F}_0(T,r)$---see \Eqn{eq:Weiss0}---which we have rewritten in a more convenient way using an integration by parts. The last line comes from the contribution $\tilde U^+\tilde V^+$ in \eqn{appeq:final}. We can conveniently rewrite the corresponding integral as
\beq
\int_0^\infty \!\!\!dx\,x^2\,{\rm Im}\left.\frac{\partial \tilde n_{\sqrt{x^2+z}-ir}}{\partial z}\right|_{z=0}=\frac{1}{2}\,\frac{\partial}{\partial r}\int_0^\infty\!\!\!dx\,x\,{\rm Re}\,\tilde n_{x-ir}.
\eeq
We then get, in terms of the polynomials $P_{2n+1}(r)$ and $P_{2n+2}(r)$ introduced in Appendix~\ref{appsec:massless},
\begin{align}
v_\infty(r)& =  -\frac{1}{3\pi^2}\Big[P_4(0)+2P_4(r)\Big]\nonumber\\
& + \frac{g^2}{2\pi^4}\Big\{\Big[P_2(r)+2P_2(0)\Big]P_2(r)+P_2'(r)P_3(r)\Big\}\,.
\end{align}
The extrema of $v_\infty(r)$ obey the following equation
\beq
0=(\pi-r)[8\pi^2(2\pi-r)r+g^2(-6\pi^2+14\pi r-7r^2)]
\eeq
which admits the solutions $r=\pi\equiv r_{\rm conf}$ and
\beq
r=\pi\left(1\pm\sqrt{\frac{8\pi^2+g^2}{8\pi^2+7g^2}}\right)\equiv r^\pm_\infty\,.
\eeq
The curvature of $v_\infty(r)$ at $r_{\rm conf}$ is $-(8\pi^2+g^2)/(24\pi^2)<0$ and the curvature at $r^\pm_\infty$ is $(8\pi^2+g^2)/(12\pi^2)>0$. There is another extremum at the boundary $r=0$ but it is a maximum since the derivative of $v_\infty(r)$ at this point is $-g^2/(4\pi)<0$. So the absolute minimum is located at $r^\pm_\infty$.

%%%%%
\section{The Polyakov loop}\label{appsec:PL}

We compute the expectation value of the traced Polyakov loop matrix field at one-loop order. We first consider the general SU($N$) case with an arbitrary representation and specialize to the case of interest in this work---namely, the fundamental representation of SU($2$)---at the end of the section. The Polyakov loop matrix field in the presence of a nonvanishing background $\bar A_0$ is defined as 
\beq
\label{appeq:popol}
L({\bf x})\equiv P\exp\left\{ig_0\int_0^\beta d\tau \left[\bar A_0+a_0(\tau,{\bf  x})\right]^{\rm bare}\right\},
\eeq
where the (bare) field $ia_0=ia_0^\kappa t^\kappa_R$ belongs to a representation $R$ of dimension $d_R$ of the SU($N$) Lie algebra and $i\bar A_0$ is restricted to the corresponding Cartan subalgebra. Here, we have emphasized that the coupling and fields appearing in the expression of the Polyakov loop matrix are the bare ones. Finally, $P$ orders these fields from left to right, with decreasing value of their time argument. The expectation value of the traced Polyakov loop is defined as
\beq
 \ell=\frac{\langle\tr\,L(\bf x)\rangle}{d_R},
\eeq
where the right-hand side is to be evaluated at $\bar A_0=\bar A_0^{\rm min}$ which means, in particular, that $\langle a_0(\tau,{\bf x})\rangle=0$.

In order to compute the loop expansion of $\ell$ in the presence of the nontrivial background field, we have to expand \Eqn{appeq:popol} in powers of $g_0$, with  $g_0\bar A_0\sim{\cal O}(1)$. For this purpose it is useful to rewrite \Eqn{appeq:popol} in the following way. Consider the time-dependent gauge transformation (written in terms of bare field and coupling)
\beq
\label{appeq:gaugebckg}
 U(\tau)=\exp\left\{-i\tau g_0\bar A_0\right\},
\eeq
under which [see \Eqn{eq:transfofo}]
\beq
 \bar A_0\to U(\tau)\bar A_0 U^\dagger(\tau)+\frac i {g_0}U(\tau)\partial_\tau U^\dagger(\tau)=0
\eeq
and
\beq
 a_\mu(\tau,{\bf x})\to U(\tau)\,a_\mu(\tau,{\bf x})\,U^\dagger(\tau)\equiv a_\mu^U(\tau,{\bf x}),
\eeq
such that\footnote{In fact, applying the transformation \eqn{appeq:gaugebckg} to all the fields $\varphi\equiv(a_\mu,c,\bar c,h)$, one eliminates all explicit references to the background field; see \Eqn{eq_ginvbare}. The latter then only arises implicitly through the unusual boundary conditions of the new field variables, e.g., $\varphi^U(\beta,{\bf x})=U(\beta)\varphi^U(0,{\bf x})U^\dagger(\beta)$. For the SU($2$) theory, this amounts to $\varphi^U_0(\beta,{\bf x})=\varphi^U_0(0,{\bf x})$ for the neutral components and $\varphi^U_\pm(\beta,{\bf x})=e^{\pm i r}\varphi^U_\pm(0,{\bf x})$, with $r=\beta g\bar A_0$, for the charged components. One immediately sees that for $r=\pi$, the charged components acquire fermioniclike antiperiodic conditions. These considerations are similar to those developed in Ref.~\cite{Roberge:1986mm} for an imaginary chemical potential.} $\bar A_0+a_0(\tau,{\bf x})\to a_0^U(\tau,{\bf x})$.
Using the standard transformation law of the Wilson line under gauge transformations \cite{Svetitsky:1985ye}, we deduce that 
\beq
\label{appeq:newpopol}
L({\bf x}) =U^\dagger(\beta)\,P\exp\left\{ig_0\int_0^\beta d\tau\, [a_0^U(\tau,{\bf  x})]_{\rm bare}\right\}U(0).
\eeq
The loop expansion in the presence of the background field simply amounts to expanding the path-ordered exponential in \Eqn{appeq:newpopol} in powers of $a_0^U$.

Up to here, we have only considered bare fields and couplings. Turning to renormalized quantities, one has to carefully take into account the fact that the background and fluctuating fields $\bar A$ and $a$ renormalize differently; see \Eqn{eq:renormfactors}. With our choice of renormalization condition \eqn{eq:condgA}, we have, in terms of renormalized quantities, $g_0\bar A_0^{\rm bare}=g\bar A_0$ for the background field contribution, and  $g_0a_0^{\rm bare}=Z_g\sqrt{Z_a}g a_0$ for the fluctuating part. However, at the order of approximation considered here, i.e., ${\cal O}(g^2a_0^2)$, the renormalization factors only appear in the correction term and can thus be safely ignored. We thus directly use renormalized quantities in the rest of this section.

In order to compute $\ell$ at one-loop order, we expand the path-ordered exponential in \Eqn{appeq:newpopol} up to quadratic order in $a_0^U$. The leading- (zeroth-) order contribution is simply 
\beq
 L^{(0)}({\bf x})=U^\dagger(\beta)=\exp\left\{ig\beta\bar A_0\right\}.
\eeq
The ${\cal O}(g)$ contribution, linear in $a_0$, gives a vanishing contribution to $\ell$, because of the condition $\langle a_0\rangle=0$. Finally, the ${\cal O}(g^2)$ (one-loop) contribution, quadratic in $a_0$, yields, for the trace,
\beq\label{appeq:F8}
{\rm tr}\,L^{(1)}({\bf  x})=(ig)^2\!\!\int_0^\beta \!\!\!d\sigma\!\!\int_0^{\sigma} \!\!\!d\tau\, {\rm tr}\!\left\{U^\dagger(\beta)\,a_0^U(\sigma,{\bf  x})\,a_0^U(\tau,{\bf  x})\right\}\!.
\eeq
Writing $\int_0^\beta d\sigma\int_0^\sigma d\tau=\int_0^\beta d\tau\int_\tau^\beta d\sigma$, renaming $\tau\to\sigma$ and $\sigma\to\tau+\beta$, and using the periodicity property 
\beq
 U^\dagger(\beta)a_0^U(\tau+\beta)=a_0^U(\tau)U^\dagger(\beta),
\eeq
\Eqn{appeq:F8} rewrites as
\beq\label{appeq:F10}
{\rm tr}\,L^{(1)}({\bf  x})\!=\!(ig)^2\!\!\int_0^\beta \!\!\!d\sigma\!\!\int_{\sigma-\beta}^{0} \!\!\!d\tau\, {\rm tr}\!\left\{U^\dagger(\beta)\,a_0^U(\sigma,{\bf  x})\,a_0^U(\tau,{\bf  x})\right\}\!.
\eeq
The half sum of \eqn{appeq:F8} and \eqn{appeq:F10} thus yields, upon renaming $\tau\to\tau+\sigma-\beta$,
\begin{align}
&{\rm tr}\,L^{(1)}({\bf  x})\nn
&\,\,=\frac{(ig)^2}{2}\!\!\int_0^\beta \!\!d\tau\!\int_0^\beta \!\!d\sigma\, {\rm tr}\!\left\{U^\dagger(\beta)\,a_0^U(\tau+\sigma,{\bf  x})\,a_0^U(\sigma,{\bf  x})\right\}\!.
\end{align}
We thus get, for the corresponding ${\cal O}(g^2)$ contribution to the average of the traced Polyakov loop,
\begin{align}
\label{appeq:PL}
\ell^{(1)}&=\frac{\langle\tr L^{(1)}(\bf x)\rangle}{d_R}\nn
&=-\frac{g^2\beta}{2d_R}\int_0^\beta \!\!d\tau \,G^{\kappa\lambda}_{00}(\tau,{\bf  0})\,\tr\!\left\{ U^\dagger(\beta-\tau)\,t^\kappa_R\,U^\dagger(\tau)\,t^\lambda_R\right\}\!,
\end{align}
where  $G^{\kappa\lambda}_{\mu\nu}(\tau-\tau',{\bf x}-{\bf x}')=\langle a_\mu^\kappa(\tau,{\bf  x})a_\nu^\lambda(\tau',{\bf  x}')\rangle$. Equation~\eqn{appeq:PL} holds for an arbitrary group SU($N$) and for any representation.

We now specify to the fundamental representation of the SU(2) group: $d_R=N=2$ and $t^\kappa_R=t^\kappa=\sigma^\kappa/2$, with $\sigma^{0}=\sigma^3$ and $\sigma^\pm=(\sigma^1\pm i\sigma^2)/\sqrt{2}$, where $\sigma^i$ are the Pauli matrices. Introducing standard up and down states \mbox{$|s=\pm\rangle$} such that $\sigma^0|s\rangle=s|s\rangle$, $\sigma^s|s\rangle=0$, and $\sigma^s|-s\rangle=\sqrt{2}|s\rangle$, we have
\beq
\bar A_0=\frac{\bar A_0^3}{2}\sum_{s=\pm}s |s\rangle\langle s|
\eeq
and thus
\beq
U^\dagger(\tau)=\sum_{s=\pm} e^{is\tau g\bar A_0^3/2} |s\rangle\langle s|\,.
\eeq
It follows that
\begin{align}
\label{appeq:traceeee}
&\tr\!\left\{ U^\dagger(\beta-\tau)\,t^\kappa\,U^\dagger(\tau)\,t^\lambda\right\}\nn
&=\sum_{s,r}e^{i\left[s(\beta-\tau)+r\tau\right]g\bar A_0^3/2}\langle s|t^\kappa|r\rangle\langle r|t^\lambda|s\rangle,
\end{align}
which is trivially evaluated. For the cases of interest in \Eqn{appeq:PL}, the trace \eqn{appeq:traceeee} gives
\begin{align}
 \frac{\cos(\beta g\bar A_0^3/2)}{2}\quad&{\rm if} \quad\kappa=\lambda=0,\\
 \frac{e^{i\kappa (\beta-2\tau)g\bar A_0^3/2}}{2}\quad&{\rm if} \quad\kappa=-\lambda=\pm.
\end{align}
We then obtain, with $r=\beta g\bar A_0^3$,
\begin{align}
\ell^{(1)} & = -\frac{g^2\beta}{8}\cos\left(\frac{r}{2}\right)\int_0^\beta d\tau \,G_{00}^{00}(\tau,{\bf  0})\nonumber\\
& - \frac{g^2\beta}{4}\int_0^\beta d\tau \,G_{00}^{+-}(\tau,{\bf  0})e^{i(r/2)(1-2\tau/\beta)},
\end{align}
where we have used that $G_{00}^{-+}(\beta-\tau,{\bf 0})=G_{00}^{+-}(\tau,{\bf 0})$ and the change of variables $\tau\to\beta-\tau$. Equivalently, with $\hat r=rT$,
\begin{align}
\ell^{(1)} & = -\frac{g^2\beta}{8}\cos\left(r\over2\right)\int_{\bf{k}} G_{00}^{0}(0,k)\nonumber\\
\label{appeq:ijsbndfv}
& -  \frac{g^2\beta}{2}\sin\left(\frac{r}{2}\right)\int_K\frac{G_{00}^+(\omega_n,k)}{\omega_n+\hat r},
\end{align}
in terms of the propagators in Fourier space. With the conventions of Sec.~\ref{sec:Frules} for the propagators, we have
\beq
 G^{\kappa\lambda}_{\mu\nu}(\tau,{\bf x})=\delta^{\kappa,-\lambda}\int_K\,e^{-i\omega_n\tau-i{\bf k}\cdot{\bf x}}\,G^\kappa_{\mu\nu}(\omega_n,k),
\eeq
where $G^\kappa_{\mu\nu}(\omega_n,k)=G^\kappa_{\mu\nu}(K)$ is defined in \eqn{eq:glupropag}.

Now, using Eqs. \eqn{eq:glupropag} and \eqn{eq:propalpha}, we have
\beq
G_{00}^0(0,k) =\frac{1}{k^2+m^2}
\eeq
and
\beq
G_{00}^+(\omega_n,k)  = \frac{k^2}{(\omega_n+\hat r)^2+k^2}\frac{1}{(\omega_n+\hat r)^2+k^2+m^2},
\eeq
and the evaluation of the Matsubara sum in \eqn{appeq:ijsbndfv} is straightforward. We obtain
\begin{align}
\int_K\frac{G_{00}^+(\omega_n,k)}{\omega_n+\hat r}&=\frac{1}{2}{\rm cotan}\left(\frac{r}{2}\right)\int_{\bf k}\frac{1}{k^2+m^2}\nn
&+\int_{\bf k} \,\frac{k^2}{m^2}\,{\rm Im}\,\left(\frac{n_{\varepsilon_k-i\hat r}}{\varepsilon^2_k}-\frac{n_{k-i\hat r}}{k^2}\right)\!.
\end{align}
Using dimensional regularization and \eqn{appeq:imn}, we arrive at the following expression for the relative ${\cal O}(g^2)$ correction to the average of the traced Polyakov loop
\begin{widetext}
\beq
\frac{\ell^{(1)}}{\cos(r/2)} =\frac{3g^2\beta m}{32\pi}+\frac{g^2\beta}{4\pi^2}\sin^2\left(\frac{r}{2}\right)\int_0^\infty dk\,\frac{k^2}{m^2}\,\left[\frac{1}{\cosh(\beta k)-\cos r}-\frac{k^2}{\varepsilon_k^2}\frac{1}{\cosh(\beta\varepsilon_k)-\cos r}\right].
\eeq
It is positive and we check that the Polyakov loop at one-loop order vanishes if and only if $r=\pi\,({\rm mod}\,2\pi)$. Moreover, in the limit $T\to\infty$ the average of the traced Polyakov loop reads
\bea
\ell^{\rm 1loop}_\infty & = & \cos\left(\frac{r_\infty }{2}\right)-\frac{g^2}{4\pi^2}\sin\left(\frac{r_\infty }{2}\right)\int_0^\infty dx\,x^4\frac{\partial}{\partial z}\left(\frac{{\rm Im}\,\tilde n_{\sqrt{x^2+z}-ir_\infty}}{x^2+z}\right)_{z=0}\nonumber\\
& = & \cos\left(\frac{r_\infty }{2}\right)+\frac{g^2}{4\pi^2}\sin\left(\frac{r_\infty }{2}\right)\Big[P_1(r_\infty)-\frac{1}{2}P'_2(r_\infty)\Big]\nonumber\\
& = & \cos\left(\frac{r_\infty }{2}\right)+\frac{3g^2}{16\pi^2}(\pi-r_\infty)\sin\left(\frac{r_\infty }{2}\right)\,,
\eea
where $\tilde n$ and $r_\infty$ were introduced in Appendix~\ref{appsec:rinf}, $P_{1,2}(r)$ given in Appendix~\ref{appsec:massless}, and we have used an integration by parts.
\end{widetext}

%%%%%
\section{Proof that $\ell=0$ at $r=\pi$}
\label{appsec:proof}

We show that the expectation value of the traced Polyakov loop vanishes for a
background field $r=\pi$ in the SU($2$) theory. We consider bare quantities throughout this section. It proves useful to introduce the following generalization of \Eqn{eq:ploop}:
\beq
\label{appeq:ploopext}
 \ell(T,r)=\frac{1}{2}\tr\left< P\exp\left\{ig_0\int_0^\beta \!\!d\tau (\bar A_0+a_0)\right\}\right>_{\langle a\rangle=0},
\eeq
where the left-hand side denotes the average with the gauge-fixed action (\ref{eq_gf}), evaluated at an arbitrary background $r=\beta g_0\bar A_0^3$, with the constraint  $\langle a(x)\rangle=0$ (this can be imposed by an appropriate source term). The physical expectation value of the traced Polyakov loop is obtained as $\ell(T)=\ell(T,r_{\rm min}(T))$.

Now, we make use of the various transformations already invoked in Sec.~\ref{sec:IIIA} to discuss the symmetry
properties of the potential $V(T,r)$. We first perform a global rotation of angle $\pi$ around the axis 1 in
color space, both for the integration variables $(a,c,\bar c,ih)$ under the path integral and for the background field. For the latter, this amounts to $r\to-r$. The integration measure---including the (gauge-fixed) action and the constraint $\langle a(x)\rangle=0$---and the Polyakov loop in the brackets of \eqn{appeq:ploopext} being invariant under global color rotations, we conclude that
\beq
 \ell(T,r)=\ell(T,-r).
\eeq

Next, we perform a gauge transformation of the form \eqn{eq_ginvbare} with a local SU($2$)
matrix $U(\tau)=\exp(-i\tau\sigma_3/\beta)$, with $\sigma_i$ the Pauli matrices. This matrix is periodic in the time direction up to an element of the center, i.e., $U(\beta)=-U(0)$. This produces a shift of the background field $r\to r+2\pi$ but leaves the integration measure, again including the (gauge-fixed) action and the constraint $\langle a(x)\rangle=0$, invariant, while the Polyakov loop gets multiplied by a phase $e^{i\pi}=-1$. We thus get
\beq
 \ell(T,r)=-\ell(T,r+2\pi).
\eeq

Combining the above identities, we get
\beq
 \ell(T,\pi-r)=-\ell(T,\pi+r),
\eeq
from which we conclude that $\ell(T,\pi)=0$. It follows that the physical expectation value of the Polyakov loop $\ell(T)=0$ if the minimum of the background field potential is $r_{\rm min}(T)=\pi$. This proof can easily be generalized, along similar lines to SU($N$).

\end{document}